\documentclass{emulateapj}
\usepackage{graphics}

\def\cm-3{\,{\rm cm^{-3}}}

\def\kpc-3{\,{\rm kpc^{-3}}}
\def\myr-1{\,{\rm Myr^{-1}}}

\def\kpc{\,{\rm kpc}}

\def\t0{$t_{\rm{cool,0}}\ $}

\usepackage{natbib}
\usepackage{color}
\usepackage{rotating}
\usepackage{footnote}
\usepackage{datetime}
\usepackage{amsmath}
\usepackage{mathrsfs}
\usepackage[normalem]{ulem}
\begin{document}

\title{The fate of AGB wind in massive Galaxies and the ICM}

\author{Yuan Li\altaffilmark{1,3} , Greg L. Bryan\altaffilmark{1,2},  Eliot Quataert\altaffilmark{3}} 

%\altaffiltext{$\dagger$}{yli@flatironinstitute.org}
\altaffiltext{1}{Center for Computational Astrophysics, Flatiron Institute, 162 5th Ave, New York, NY 10010, USA; yli@flatironinstitute.org}
\altaffiltext{2}{Department of Astronomy, Columbia University, 550 W 120th Street, New York, NY 10027, USA} 
\altaffiltext{3}{Department of Astronomy, and Theoretical Astrophysics Center, University of California, Berkeley, CA 94720, USA}

\begin{abstract}
{Asymptotic Giant Branch (AGB) winds from evolved stars not only provide a non-trivial amount of mass and energy return, but also produce dust grains in massive elliptical galaxies. Due to the fast stellar velocity and the high ambient temperature, the wind is thought to form a comet-like tail, similar to Mira in the Local Bubble. Many massive elliptical galaxies and cluster central galaxies host extended dusty cold filaments. The fate of the cold dusty stellar wind and its relation to cold filaments are not well understood. In this work, we carry out both analytical and numerical studies of the interaction between an AGB wind and the surrounding hot gas. We find that the cooling time of the tail is inversely proportional to the ambient pressure. In the absence of cooling, or in low pressure environments (e.g., the outskirts of elliptical galaxies), AGB winds are quickly mixed into the hot gas, and all the AGB winds have similar appearance and head-to-tail ratio. In high pressure environments, such as the Local Bubble and the central regions of massive elliptical galaxies, some of the gas in the mixing layer between the stellar wind and the surrounding hot gas can cool efficiently and cause the tail to become longer. Our simulated tail of Mira itself has similar length and velocity to that observed, and appears similar to the simulated AGB tail in the central regions of massive galaxies. We speculate that instead of thermal instability, the induced condensation at the mixing layer of AGB winds may be the origin of cold filaments in massive galaxies and galaxy clusters. This naturally explains the existence of dust and PAH in the filaments. }
\end{abstract}

\section{Introduction}
\label{sec:intro} 

Old stars return a significant fraction of their original mass to their surrounding medium during the AGB phase. In massive elliptical galaxies, the total mass loss rate can be as high as a few solar masses per year, exceeding the total star formation rate in many systems \citep{Leitner2011, Voit2011, McDonald2018}. Most early-type galaxies have very low-level or no star formation activity, and therefore very low core-collapse supernova rate. AGB stars are not only an important source of mass, but are also responsible for producing most, if not all, of the dust grains \citep{Ferrarotti2006}. 

A significant fraction of cool-core galaxy clusters and early-type galaxies harbor extended multiphase gas in their centers, which can be often observed in emission lines \citep{Edge2001, Salome2003, McDonald10, Werner2014, Pandya2017}. The cold gas is generally thought to have condensed out of the hot intra-cluster medium (ICM) due to thermal instability \citep{McCourt12, Sharma2012} triggered by AGN uplifting \citep{PII, Voit2017} and turbulence \citep{Voit2018}. In massive galaxies and clusters with virial temperatures above a few $10^6$ K, dust sputtering time is extremely short, and the hot gas in these systems should be dust-free. If condensation from dust-free hot gas is the origin of the filamentary cold gas, then the filaments should also be dust free. However, dust has long been observed to exist in these line-emitting filaments \citep{Sparks1989, Goudfrooij1994}. Moreover, \citet{Donahue2011} show that the dust grains there are similar to those of normal star forming galaxies, suggesting that the grains have been shielded from the hot gas since they were produced at the surface of AGB stars. 

The fate of AGB winds in massive galaxies is not well understood. Because the stars are typically moving at 300-400 km/s in massive elliptical galaxies, the stellar wind is pushed to the trailing side of the star and forms a head-tail structure. Analytical work by \citet{Mathews1990} suggests that cold AGB winds should quickly assimilate into the hot phase as a result of Rayleigh-Taylor and Kelvin-Helmholtz instabilities, along with thermal conduction. \citet{Parriott2008} carry out 2D hydrodynamical simulations of an AGB wind interacting with a hot ISM wind in environments characteristic of the outer parts of low mass elliptical galaxies. Their simulations show that up to $\sim 20\%$ of the cold wind can survive rather long, $\sim 20$ pc from the star. So far, there has not been any 3D numerical work on AGB wind in massive galaxies. \citet{Voit2011} suggests that cold AGB winds should survive longer in higher pressure environments, which has not been explored in previous numerical work either.

It is impossible to directly observe individual AGB winds in any external massive galaxies. Most of the nearby AGB stars in our Milky Way are moving in a very different environment, and at a much lower velocity, so the interaction between the winds and the ISM is very different. One exception is Mira the Wonderful. Mira is a binary system in the Local Bubble \citep{Lallement2003}, where Mira A is a luminous variable AGB star, and Mira B is generally thought to be a white dwarf. Unlike the majority of the disk stars in the Milky Way, Mira has a very large space velocity of $\sim 130$ km/s \citep{Evans1967}. \citet{Martin2007} first discovered Mira's comet-like tail using GALEX. The UV emission is thought to come from collisionally ionized $H_2$ at the interface of the cool AGB wind from Mira A and the surrounding hot gas. The formation of the tail is a result of the interaction between the AGB wind and the fast moving ISM (from the star's perspective). Given Mira's large velocity and its environment, \citet{Conroy2015} suggest that its configuration is common in the central regions of early-type galaxies. 

In this paper, we study the interaction between AGB winds and the surrounding hot gas using 3D hydro simulations. We perform three sets of simulations, including Mira, an AGB star in an environment that resembles the outer regions of early-type galaxies, and a star in the central regions. Each set of simulations consists of an adiabatic run and a comparison run with radiative cooling. We focus on the early AGB phase and assume a constant stellar mass loss rate. Most of the mass loss happens at the end of the AGB phase, during the formation of a planetary nebula. However, the early AGB phase lasts much longer, allowing the AGB star to interact with more ISM. We leave the interaction between the planetary nebula and the hot ISM to future studies. 

The paper is structured as follows: in Section~\ref{sec:analytic}, we present analytical calculations of the properties of the tail of a fast moving AGB star, including the morphology of the tail, and how its cooling time depends on the environment; in Section~\ref{sec:method}, we describe the simulation setup and the simulations performed; in Section~\ref{sec:results}, we present the main results of the simulations; in Section~\ref{sec:discussions}, we compare our simulated Mira's tail with the observations, discuss the implication of this work for the formation of dusty cold filaments in massive galaxies and clusters, compare our simulations with previous works, and discuss our limitations and caveats. We conclude this work in Section~\ref{sec:conclusions}.

\section{Analytic work} \label{sec:analytic}
\subsection{tail velocity, tail length and tail size}
In this section, we calculate the properties of the tail of a Mira-like star, including its velocity and length. The stellar wind shocks when its ram pressure equals the ram pressure of the ISM (this is also roughly equal to the thermal pressure of the ambient gas if the velocity of the star $v_\star$ is the same as $\sigma$ of the gas). Since the velocity of the stellar wind $v_w$ is typically much smaller than $v_\star$, we have:

\begin{equation}
\rho_w v^2_w \approx \rho_{ISM}v^2_\star \, ,
\end{equation}
where $\rho_w$ is the density of stellar wind at the contact discontinuity and $\rho_{ISM}$ is the density of the ISM. Before the stellar wind reaches the standoff radius, it is in free expansion with a constant velocity $v_w$. At the standoff radius $r_s$, the mass flux of the stellar wind (this is also the stellar mass loss rate) is

\begin{equation}
\dot{M_w}=\rho_w v_w \times 4\pi r^2_s\, .
\end{equation}

Thus the standoff radius is

\begin{equation}
r_s \approx \frac{1}{v_\star}\sqrt{\frac{\dot{M_w}v_w}{4\pi \rho_{ISM}}}.
\end{equation}

After colliding with the ISM, the stellar wind is swept to the trailing side and forms a tail. Due to mass conservation, the mass flux of the tail is roughly equal to the stellar mass loss rate (ignoring the contribution from the ISM, which is smaller by $\sim v_w/v_\star$): $\dot{M_t}\approx \dot{M_w}$. Momentum conservation gives:

\begin{equation}\label{eq:momentum}
\dot{M_t}v_t \approx \rho_t v^2_t \times \pi r^2_s \approx \rho_{ISM}v^2_\star \times \pi r^2_s \approx \frac{1}{4}\dot{M_w}v_w \, .
\end{equation}

Therefore, we have 
\begin{equation}
v_t \approx \frac{1}{4}v_w \,.
\end{equation}

This means that the velocity of the tail material depends only on the velocity of the stellar wind in the star's frame. 

In the absence of cooling, the stellar wind material in the tail eventually mixes with the ISM if the mixing timescale is shorter than the acceleration timescale. The mixing time can be estimated using the cloud shredding time measured from cloud crushing simulations \citep{Scannapieco2015}. The cloud crushing time is
\begin{equation}
t_{cc} \approx \frac{r_s}{v_\star} \sqrt{\frac{\rho_t}{\rho_{ISM}}} \approx \frac{4r_s}{v_w} \, .
\end{equation}
The shredding time is usually a few times the cloud crushing time \citep{Klein1994}: $t_{sh} \approx a t_{cc}$, where $a$ can be measured from numerical simulations. The value for $a$ can vary slightly depending on the exact simulation setup \citep{Zhang2017}. Here, for simplicity, we use $a\sim 4$, which is roughly the time at which the fraction of the mass at or above 1/3 the original density of the cloud is $75\%$ for a cloud moving at transsonic velocities \citep{Scannapieco2015}.
Therefore the length of the tail is 
\begin{equation}
l_t \approx v_t t_{sh} \approx a r_s \, .
\end{equation}

One can also estimate the shredding time using
\begin{equation}
t_{sh}\approx \frac{r_s}{\sigma_t} \approx \frac{r_s}{f v_t} \, ,
\end{equation}
where $\sigma_t$ is the typical turbulent velocity in the tail, and is a fraction ($f$) of the bulk velocity of the tail $v_t$. This gives us a similar scaling relation: $r_s=f l_t$. This means that if radiative cooling does not play a role, all Mira-like stars should have a similar appearance, with a tail-to-head ratio of $a\sim4$. 

\subsection{when cooling of the tail is important}
In this section, we analyze when radiative cooling is important in the evolution of the tail. From Equation~\ref{eq:momentum}, we have the density of the tail as
\begin{equation}
\rho_t \approx16\rho_w \approx 16\rho_{ISM}(\frac{v_\star}{v_w})^2 \, .
\end{equation}
The tail material is in rough pressure equilibrium with the surrounding ISM: $P_t\sim P_{ISM}$. The cooling time of the tail is:
\begin{equation}
t_{cool}\propto \frac{P_t}{\rho^2_t \Lambda(T)} \propto \frac{P_{ISM}}{\rho^2_t\Lambda(T)}\, ,
\end{equation}
where $\Lambda(T)$ is the cooling function. Ignoring the variation in $\Lambda(T)$ (assuming small temperature variation in the tail) and assuming again that the velocity of the star $v_\star$ is the same as $\sigma$ of the gas, we have:
\begin{equation}
t_{cool}\propto \frac{\rho_{ISM}v^2_\star}{\rho^2_t} \propto \frac{1}{\rho_{ISM}}\frac{v^4_w}{v^2_\star}  \propto \frac{v^4_w}{P_{ISM}}\, .
\end{equation}
Thus the cooling time of the tail is inversely proportional to the ambient pressure. In high pressure environments, the tail is likely to stay cold for longer. Note that this is the $t_{cool}$ of the tail material, not the cooling time of the shocked ISM, or the gas in the mixing layer between the two. Although the cooling time of the mixing layer is related to $t_{cool}$ of the tail \citep{Begelman1990, GronkeOh2018}.

\section{Methodology}
\label{sec:method}
\subsection{Simulation Setup}

The simulations described in this paper are performed using the adaptive mesh code Enzo \citep{Enzo} with the Piecewise Parabolic Method (PPM) of \citet{PPM}. The radiative cooling rate of the gas is computed based on a temperature-dependent cooling function from \citet{Sarazin1987} for temperatures above $10^4$ K, and further cooling down to 300 K is based on the rates in \citet{Rosen1995}. We assume a constant half-solar metallicity for all the gas including the stellar wind and the ISM. 

The simulation box is set up similarly to a cloud crushing simulation. Our simulation domain is a rectangular box of $L_x \times L_y \times L_z $, with $L_y=L_z$. $L_x=3$ or 4 times $L_y$ depending on the simulation. The rest frame follows the center of the AGB star. The star is placed closer to one side of the box with a distance of $0.2 L_y$ from the boundary. This boundary has a constant inflow of $v_\star$ with density $\rho_{ISM}$ and temperature $T_{ISM}$. The boundary on the opposite side is outflowing, while all other boundaries are periodic. Around the AGB star, we define a wind injection sphere with a radius typically set to be about half of the standoff radius $r_s$ defined in Section~\ref{sec:analytic}. We compute the wind density at this radius based on $\dot{M}_w$ and $v_w$. As discussed previously, the stellar ejecta follows the free expansion solution within $r_s$, with a constant velocity and $\rho \sim r^{-2}$. At every simulation time-step, we set the boundary condition at the wind injection sphere such that the stellar wind flows through the sphere following the analytical free expansion solution with fixed velocity, density and temperature $T_w$. We also inject a passive tracer fluid along with the stellar wind material. Outside of the wind injection sphere, the gas has an initial uniform density $\rho_{ISM}$ and temperature $T_{ISM}$. 

In all simulations, the short sides of the box have 32 root grids, and the number of root grids on the long side scales accordingly. The maximum level of refinement is 5. We also lay out a set of nested static refined regions around the AGB star so that the region within $r_s$ is always refined to the highest level and the whole region around the AGB star is reasonably well resolved.

\subsection{Simulations Performed}

In this section, we describe the three sets of simulations that are performed and analyzed in this paper: Mira, low pressure (LP) and high pressure (HP) simulations. Each set of simulations has two runs: the adiabatic case where radiative cooling is turned off, and the cooling case. The parameters used in the simulations are listed in Table ~\ref{table:parameters}.

Mira is a low mass AGB star located in the Local Bubble. Our Mira simulation uses parameters similar to previous simulations of Mira \citep[e.g.,][]{Wareing2007, Esquivel2010}, which were obtained based on CO line observations of Mira \citep{Young1995, Ryde2000}. The only parameter that is different from what has been used in previous simulation works is the mass loss rate, for which we have adopted a lower value ($\sim 1/4$ of what is used previously). This is because the mass loss rate used in previous simulations is based on observations of the current state of Mira \citep{Ryde2000}, but our simulation follows the evolution of Mira's tail for half a million years. The mass loss rate of Mira was likely much lower at the beginning of the simulation. Given that the mass loss rate of AGB stars increases drastically with time, especially towards the end \citep{Bowen1991}, a constant rate is not a good assumption, but it is commonly used and is also used here to simplify the physics. We plan to incorporate a more realistic time-dependent mass-loss rate based on stellar evolution models \citep{MIST} in future studies. As is shown in Section~\ref{sec:analytic} though, the shape and the cooling time of the tail are both independent of $\dot{M}_w$.

Some of the previous simulations of Mira have also employed a hypothesis that Mira has only recently entered the Local Bubble \citep{Esquivel2010, Wareing2012} in order to explain some of the observed features of Mira. Since matching the observations of Mira is not the focus of this paper, we do not introduce any complicated assumptions about Mira's environment and simply place it in the Local Bubble since the beginning of the simulation. %As we show later, our simulated Mira shows good agreement with the observations, and it is possible that the assumption of Mira recently entering the Local Bubble is unnecessary.

Our LP simulations adopt the parameters used in the fiducial run in \citet{Parriott2008}, where they simulate the wind of an AGB star interacting with low density, hot ISM. The physical condition of the environment is typical of the middle to outer regions of an elliptical galaxy, and has a lower pressure than the Local Bubble. The parameters for the AGB wind are comparable to but not exactly the same as Mira. The velocity of the star (and thus the hot ISM in the simulation) is 350 km/s, which is again the typical velocity of stars in an elliptical galaxy. Our simulation has a slightly higher resolution than \citet{Parriott2008}. The main difference is that \citet{Parriott2008} is in 2D while our simulation is in 3D. We discuss the difference in the simulation results in Section~\ref{sec:previous}.

Our third set of simulations (HP) place the AGB star in an environment typical of the central region of a massive elliptical galaxy, where the gas pressure is the highest amongst all the simulations we performed. We use the same stellar velocity of 350 km/s as the LP run. 

We run all simulations for about 0.5 Myr. This is long enough for the system to reach a steady state, as we show later. Since the lifetime of a typical AGB star is about 1 Myr, running the simulation for longer is not physically meaningful either. 

\begin{table}
\caption{Simulation Parameters}\label{table:parameters}
\begin{center}
%\centering
\begin{tabular}{ | c | c | c | c |}    \hline
   & Mira & Low Pressure & High Pressure \\ \hline
$\dot{M}_w (10^{-7} \rm M_\odot/yr)$   &    0.75     &   1   &    0.3          \\ \hline
$v_w \rm (km/s)$   & 5      &    35      & 35                  \\ \hline
$v_\star \rm (km/s)$ & 125 & 350 & 350 \\ \hline
$\rm T_{ISM}$ (K) & $10^6 $&  $3\times10^6$  &  $10^7$\\ \hline
$ \rm n_{ISM} (cm^{-3})$ & $5\times10^{-2}$ & $10^{-3}$ & $10^{-2}$ \\ \hline
$r_s$\footnote{The standoff radius is not a parameter in the simulation setup, but it does affect the sizes of the static refine regions as described in Section~\ref{sec:method}.} (pc) & 0.08 & 0.5 & 0.06 \\ \hline \hline
box size (pc) & $2\times2\times8$ & $20\times20\times60$   & $2\times2\times6$ \\

    \hline 
    \end{tabular}\\ \ \\
    \end{center}
\end{table}

\section{Simulation Results}\label{sec:results}
\subsection{Comparison to Analytic Scaling}\label{sec:results1}
\subsubsection{Adiabatic Runs}\label{sec:results1a}

\begin{figure}
\begin{center}
\includegraphics[scale=0.09,trim=2.1cm 0cm 1.5cm 0cm, clip=true]{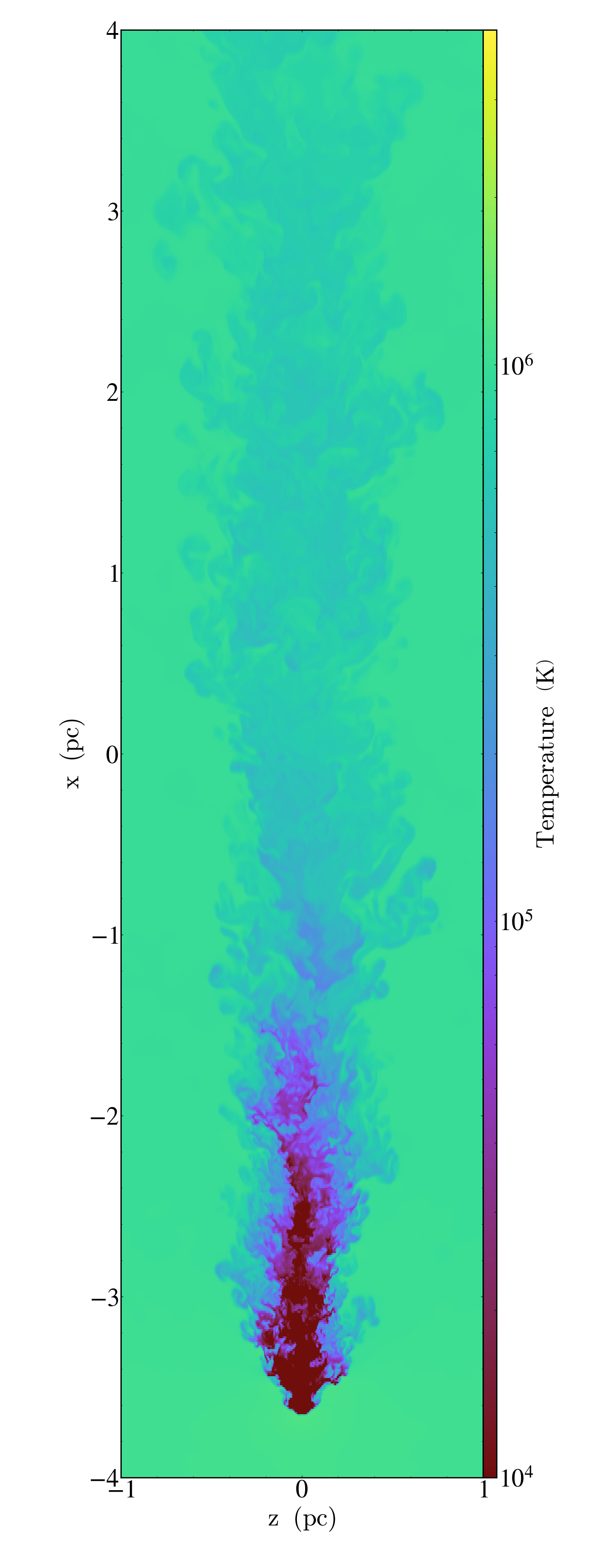}
\includegraphics[scale=0.09,trim=2.1cm 0cm 1.3cm 0cm, clip=true]{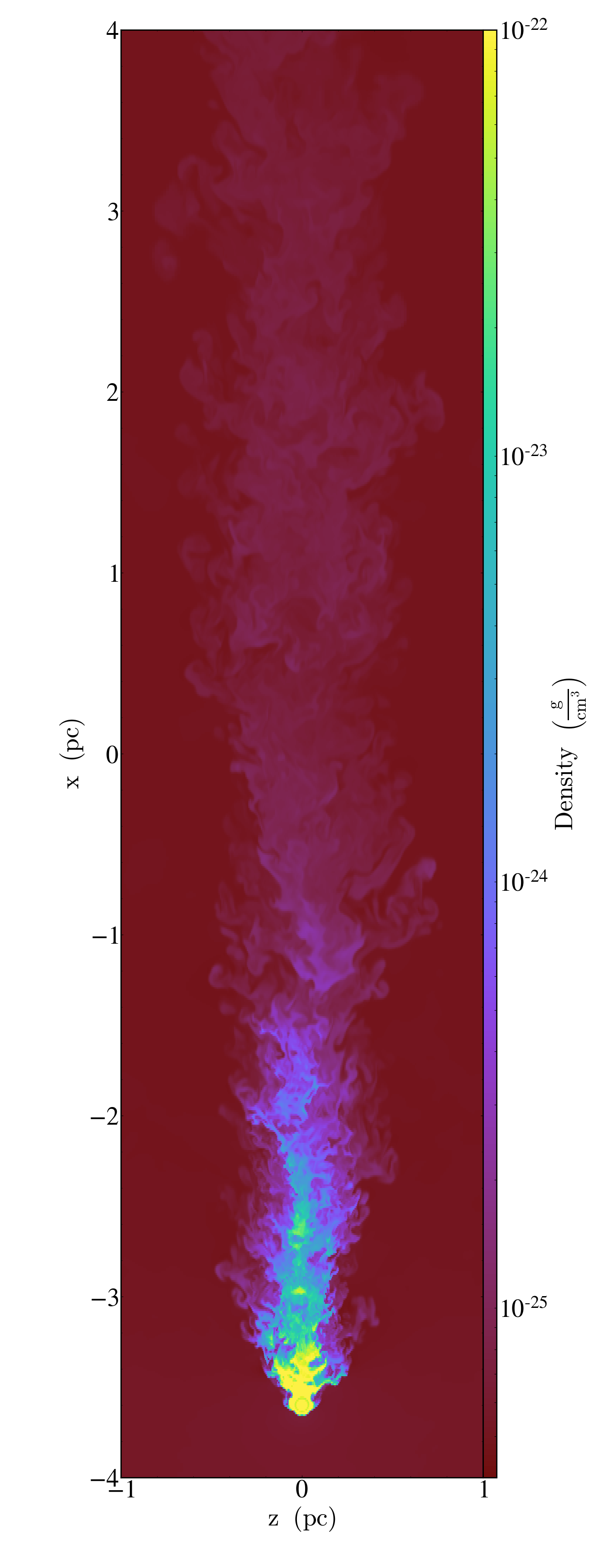}
\includegraphics[scale=0.09,trim=2.1cm 0cm 2cm 0cm, clip=true]{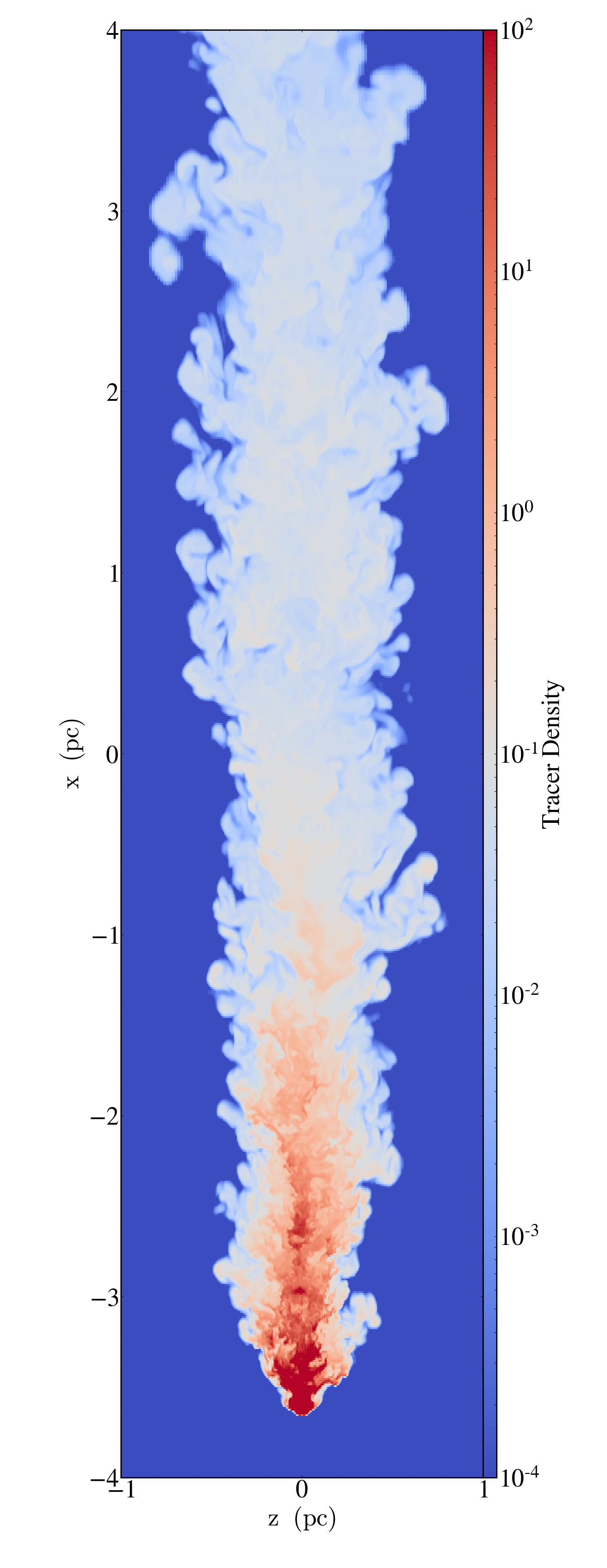}
\caption{Temperature, density and the density of tracer fluid in a slice of gas through the center of the adiabatic simulation of Mira along the direction of the ISM wind. The run with radiative cooling is shown in Figure~\ref{fig:Mira2_slice}.
\label{fig:Mira3_slice}}
\end{center}
\end{figure}

\begin{figure}
\begin{center}
\includegraphics[scale=0.4,trim=0cm 0cm 0cm 0cm, clip=true]{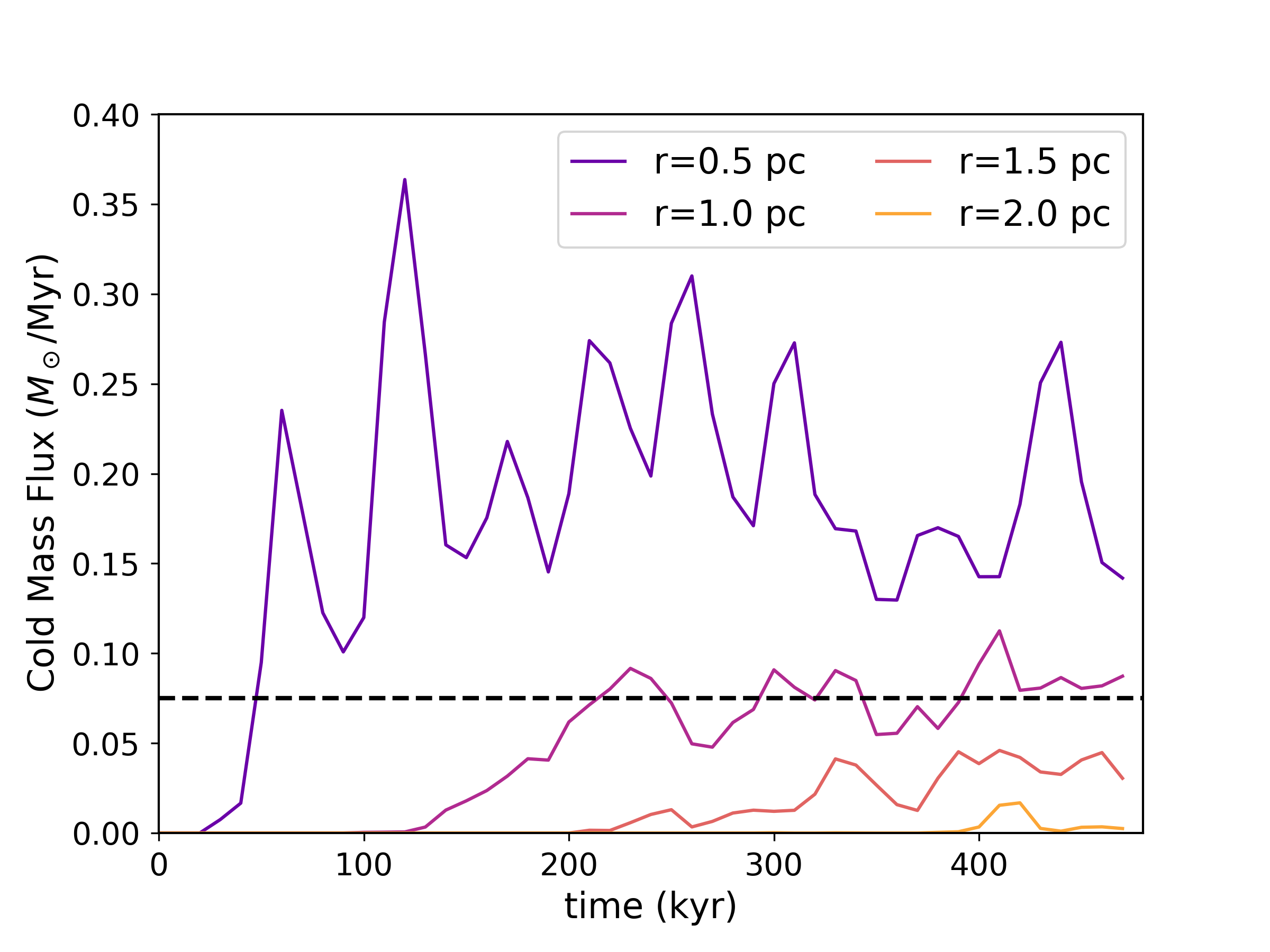}
\caption{Mass flux of cold gas (defined as $\rm T<10^4$ K) as a function of time at different distances from the star in the adiabatic simulation of Mira's tail. The black dashed line shows the cold flux ($\rm T=100$ K) from Mira. Data is sampled every $10^4 yr$.
\label{fig:Mira3_cold_flux}}
\end{center}
\end{figure}

\begin{figure}
\begin{center}
\includegraphics[scale=0.09,trim=2.1cm 0cm 1.5cm 0cm, clip=true]{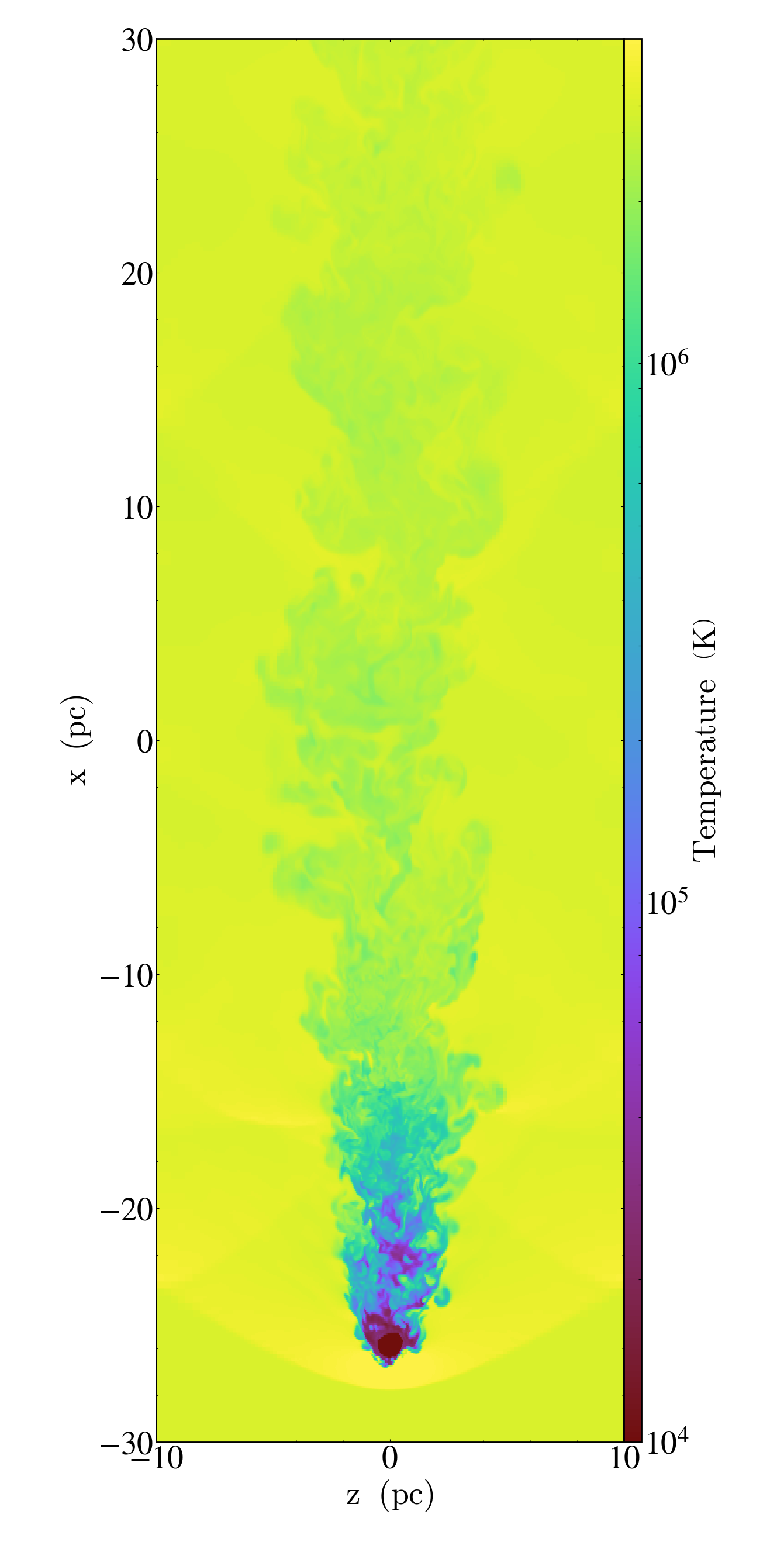}
\includegraphics[scale=0.09,trim=2.1cm 0cm 1.5cm 0cm, clip=true]{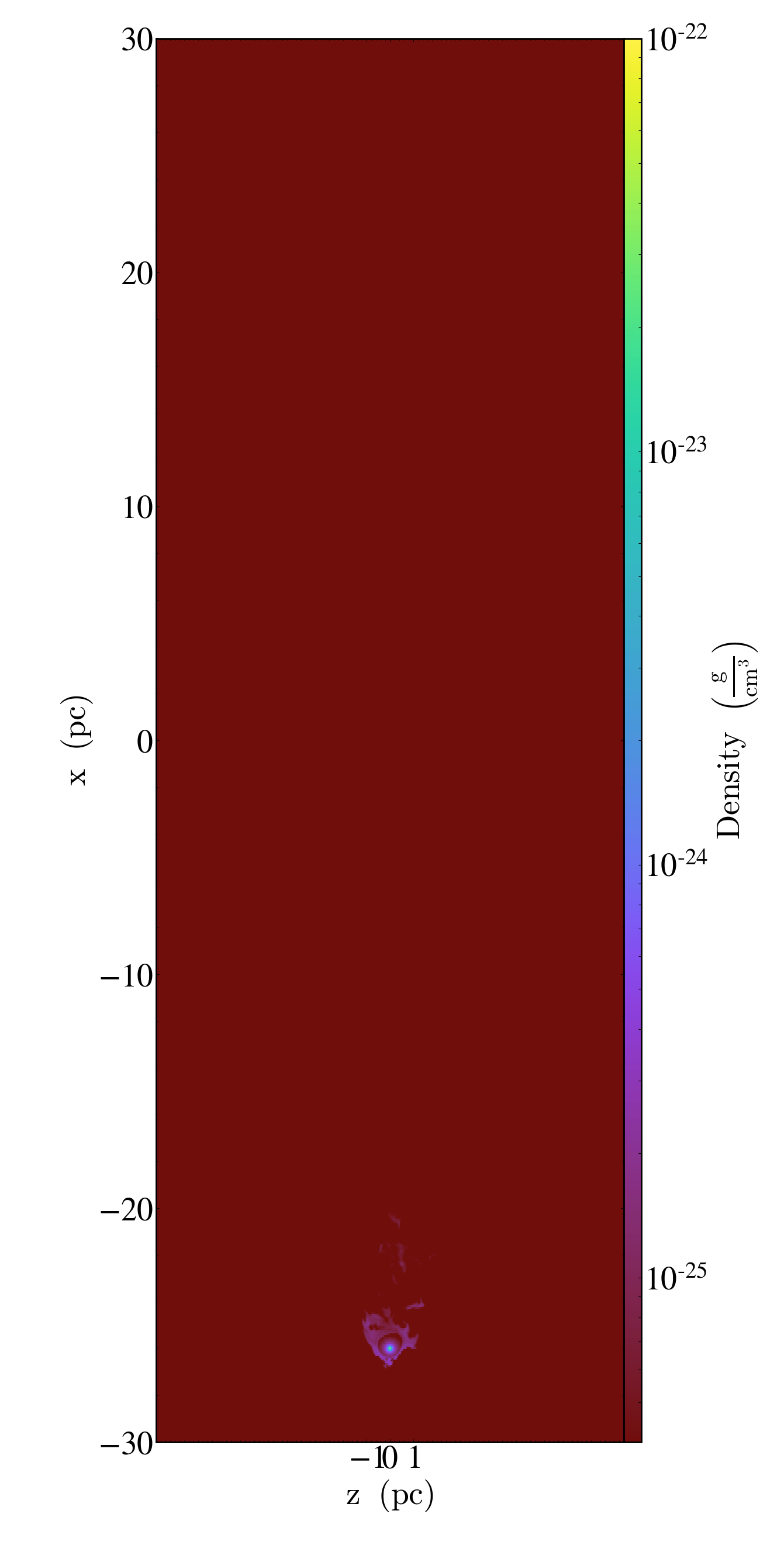}
\includegraphics[scale=0.09,trim=2.1cm 0cm 1.5cm 0cm, clip=true]{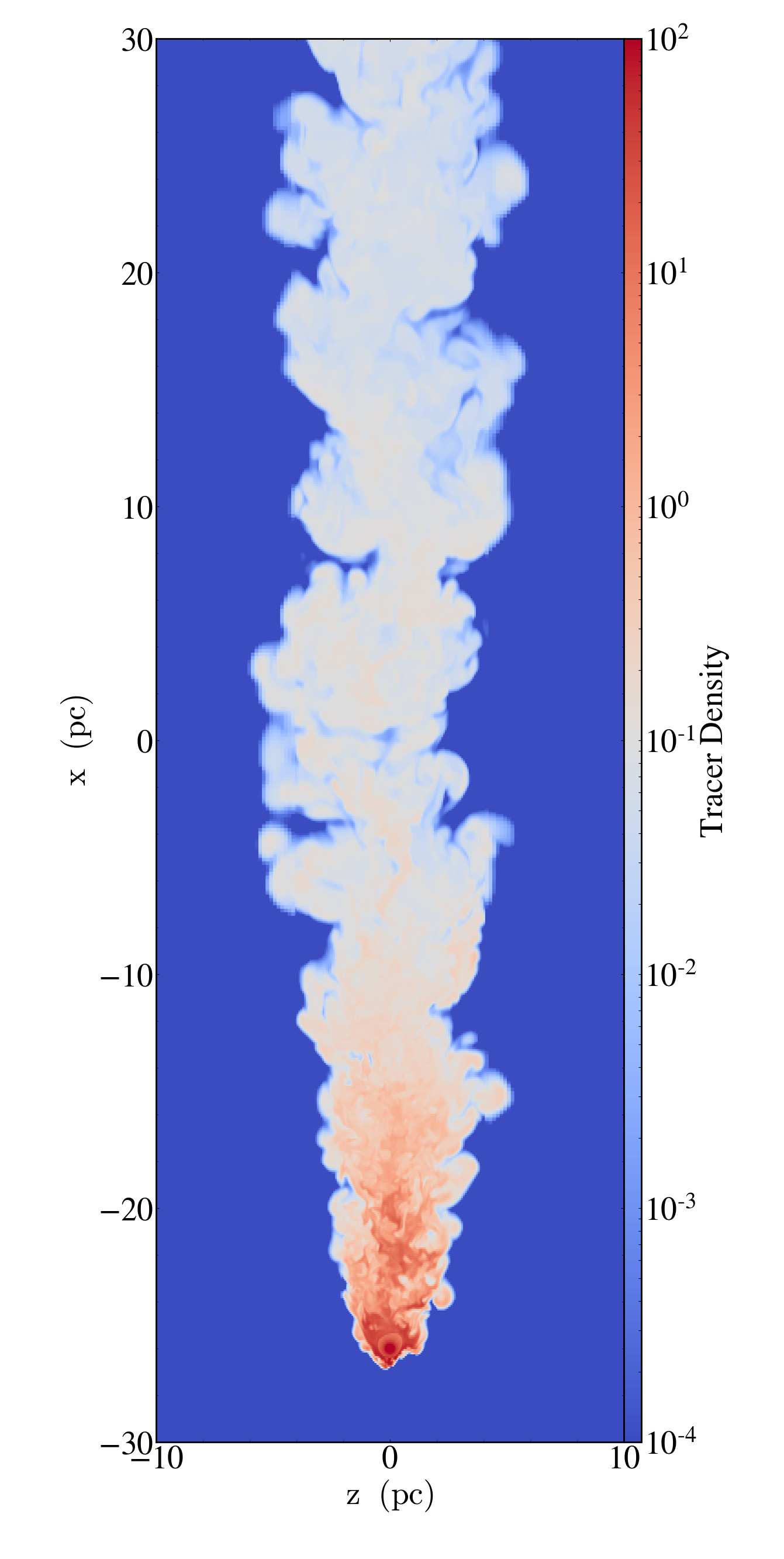} \\

\includegraphics[scale=0.09,trim=2.1cm 0cm 1.5cm 0cm, clip=true]{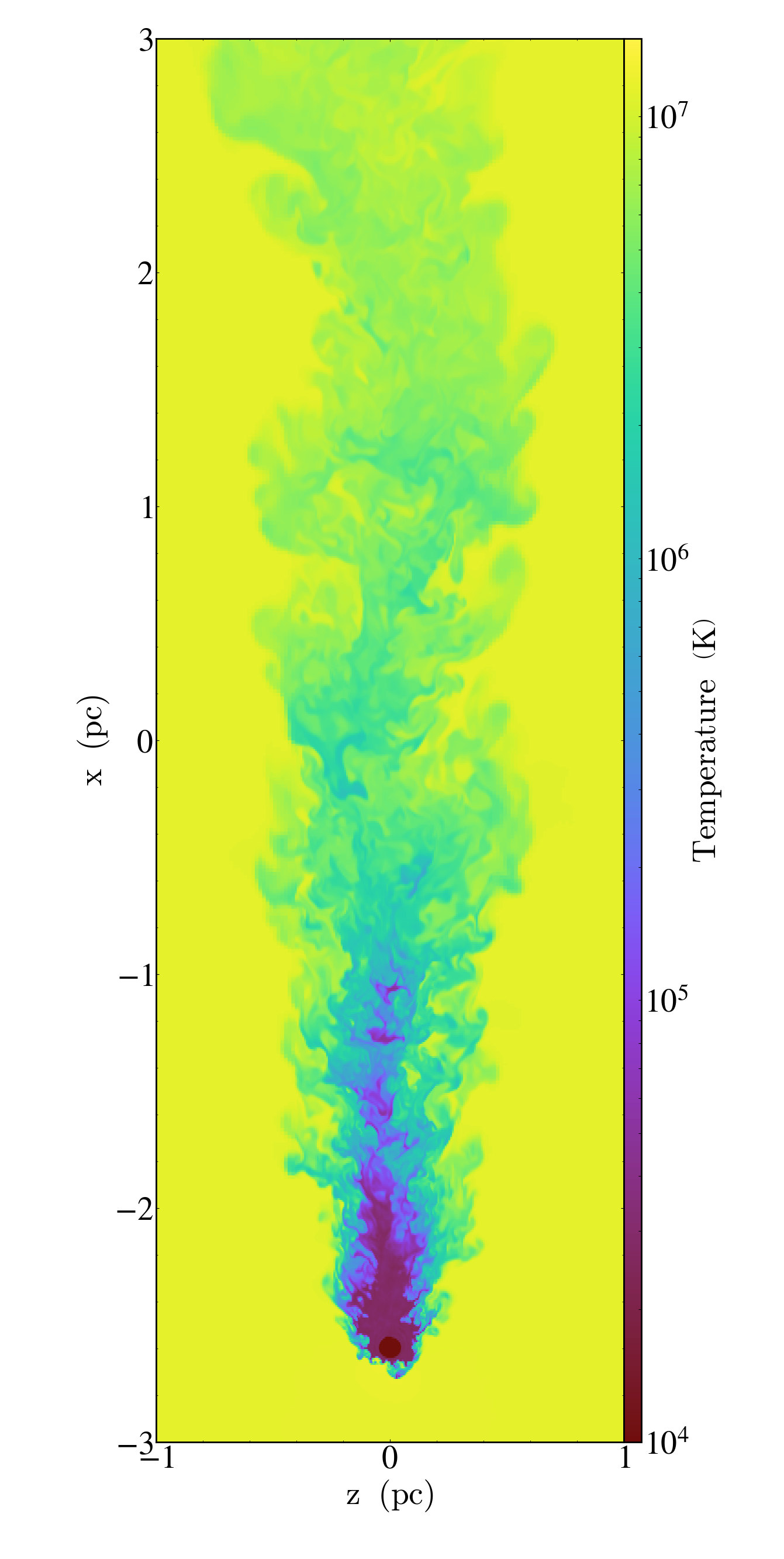}
\includegraphics[scale=0.09,trim=2.1cm 0cm 1.5cm 0cm, clip=true]{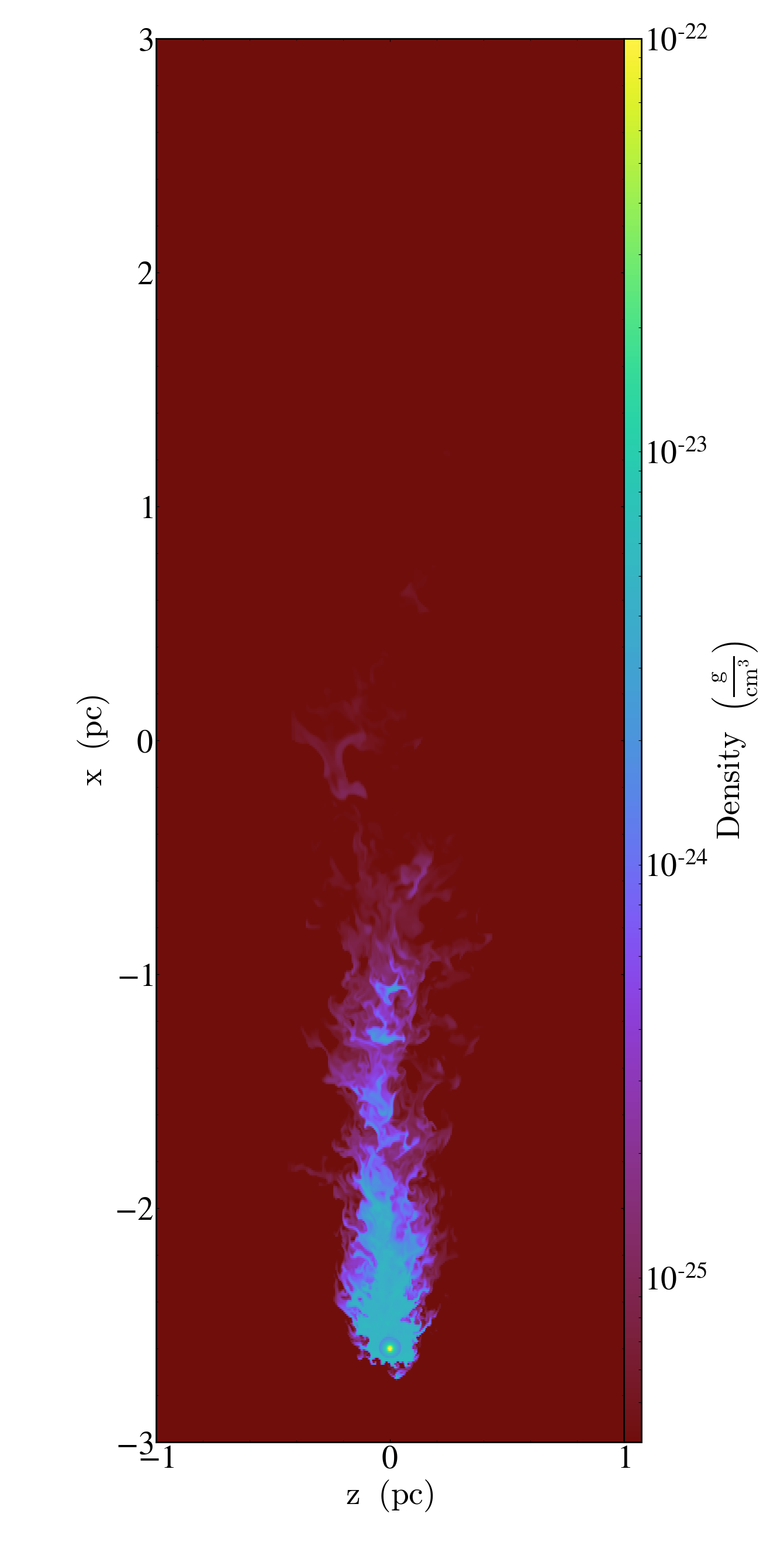}
\includegraphics[scale=0.09,trim=2.1cm 0cm 1.5cm 0cm, clip=true]{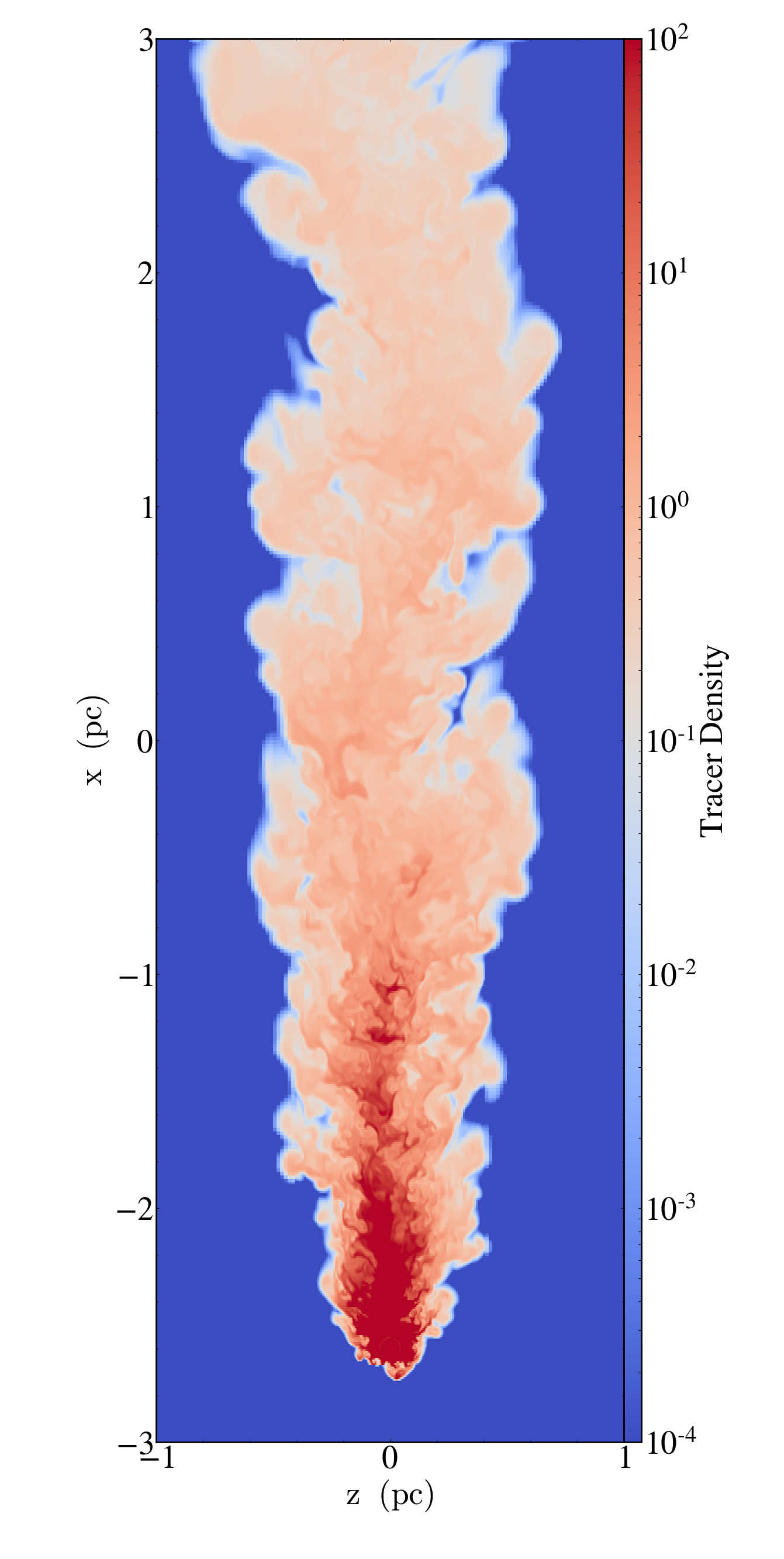}
\caption{Top panels: temperature, density and the density of tracer fluid in a slice of gas through the center of the simulation domain in the adiabatic run for the low pressure case. Bottom: temperature, density and the density of tracer fluid in a slice in the high pressure run.
\label{fig:LPHP_slice}}
\end{center}
\end{figure}

\begin{figure*}
\begin{center}
\includegraphics[scale=0.44,trim=8cm 0cm 1cm 0cm, clip=true]{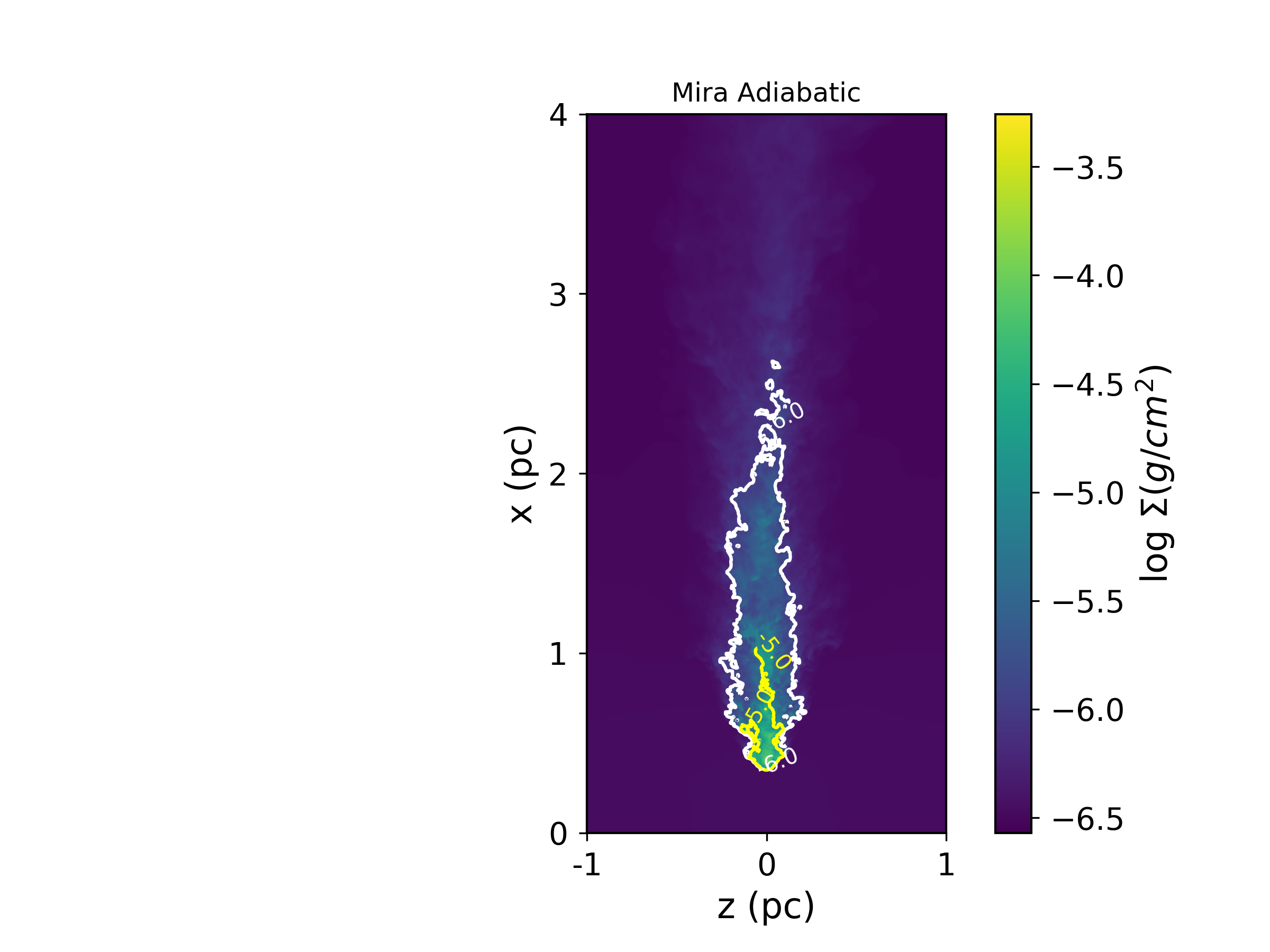}
\includegraphics[scale=0.44,trim=8cm 0cm 1cm 0cm, clip=true]{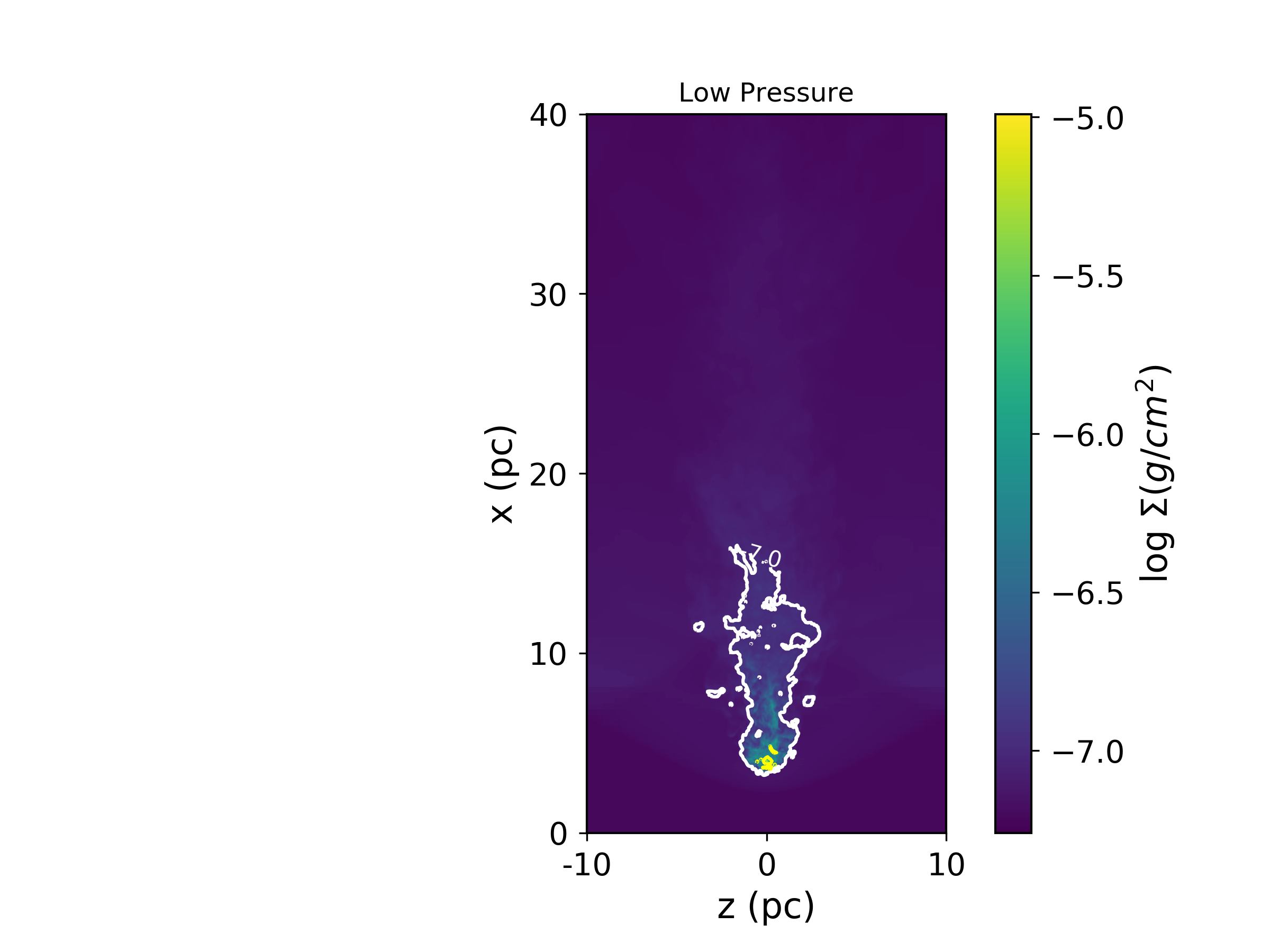}
\includegraphics[scale=0.44,trim=8cm 0cm 1cm 0cm, clip=true]{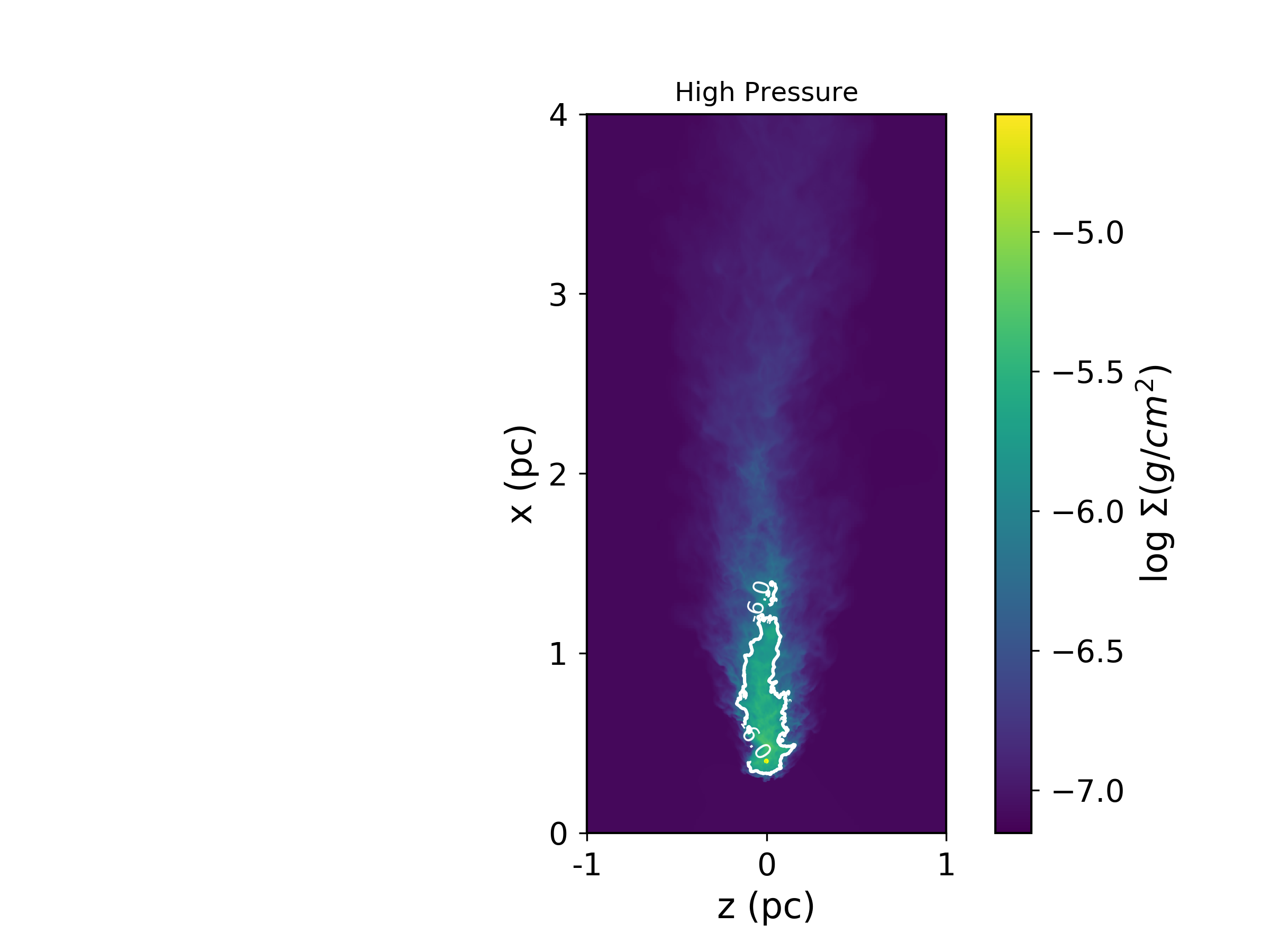}
\caption{Projections of gas density in all three adiabatic runs: Mira (left panel), LP (middle panel) and HP (right panel). Overlaid are iso-surface density contours.
\label{fig:contour}}
\end{center}
\end{figure*}

In this section, we compare our adiabatic runs with the analytical calculations, focuing on the universal shape that we found in Section~\ref{sec:analytic}. 

We start injecting stellar wind at the beginning of the simulation with uniform background density and temperature. The stellar wind quickly reaches the standoff radius $r_s$ where it becomes Rayleigh-Taylor unstable. The hot ISM blows on the stellar wind material and pushes it to the trailing side. A cold tail forms and gradually grows to a certain length when the system reaches a steady state (Figure~\ref{fig:Mira3_slice}). The cold tail is highly turbulent and the tail material is subject to Kelvin-Helmholtz instabilities due to the velocity shear between the cold tail and the ISM. As a result, the amount of cold gas decreases as a function of distance from the star. In the absence of cooling, the cold tail cannot stay cold forever and eventually fully mixes with the hot ISM at a certain distance from the star. In the steady state, the supply of cold material from the AGB star and the loss due to turbulent mixing reach a balance, and the cold tail in this steady state has a characteristic shape, as discussed in Section~\ref{sec:analytic}. 

Figure~\ref{fig:Mira3_cold_flux} shows the mass flux of cold gas (defined as T$< 2 \times 10^4$ K) through consecutive planes along the tail perpendicular to the ISM wind direction. The cold mass flux at all distances from the AGB star shows an initial increase and then settles to a quasi-steady value (though still showing large temporal variations). The steady state is reached earlier for smaller distances, as one would expect. By the end of the simulation, the flux at 2pc, the largest distance where cold gas is seen, has reached the steady state. Similarly, we have verified that all the other simulations performed have reached a steady state by 0.5 Myr, and all of our discussions are focused on the steady state unless specified otherwise.

Figure~\ref{fig:Mira3_slice} shows the temperature, density and ``color'' of a slice of gas through the center of the simulation domain along the direction of the ISM wind of the Mira simulation without radiative cooling. The snapshot is taken at $t=0.45$ Myr, during the steady state described previously. We will not specify the exact time of the snap shot throughout the rest of the paper. Any snapshot and the corresponding discussion should be treated as a typical snapshot of the steady state of the system. The cold AGB wind material forms a head-tail structure similar to the observations of Mira, but the cold tail here is only 2 pc long, about half the length of the observed Mira's tail. As we discuss in Section~\ref{sec:results1b}, the tail becomes longer when radiative cooling is included in the simulation. In the adiabatic run, after 2 pc, the AGB wind is well mixed with the hot ISM of the Local Bubble. The tail is highly turbulent. The cold tail narrows at larger distances from the star, but the tail material spreads out wider. 

Figure~\ref{fig:LPHP_slice} shows the slices for the low pressure (top panel) and high pressure (bottom panel) runs. The overall structures of the AGB wind in all three cases are similar. The wind material of in the LP run spreads out to a much larger area because of the high $v_w$ and the low ambient pressure. In all cases, the length of the cold tail appears to be a few times the width of the head, consistent with our analytical calculation in Section~\ref{sec:analytic}. 

To provide more quantitive measurements of the heat-tail structure, we make iso-surface density contours of the three cases, shown in Figure~\ref{fig:contour}. We measure the width and length of the iso-surface density contour with $\rm \Sigma=10^{-6} g/cm^2$ for Mira and the high pressure run, and use $\rm \Sigma=10^{-7} g/cm^2$ for the low pressure case. The measured head widths and tail lengths for Mira, LP and HP cases are 0.4 pc by 1.8 pc, 3 pc by 12 pc, and 0.2 pc by 0.7 pc, respectively. In all cases, the tail length to head width ratio is close to 4, consistent with what is found in Section~\ref{sec:analytic}. The shape of the contour is not very sensitive to the exact choice of the surface density cut. As the first panel of Figure~\ref{fig:contour} shows, in the case of Mira, the $\rm \Sigma=10^{-5} g/cm^2$ iso-surface density contour has a very similar shape as the $\rm \Sigma=10^{-6} g/cm^2$ one. The head width and tail length are 0.1 pc and 0.6 pc if we use a density cut of $\rm \Sigma=10^{-5} g/cm^2$. We have also experimented using iso-surface density contour of the tracer fluid, and found similar results. As Figure~\ref{fig:Mira3_slice} and Figure~\ref{fig:LPHP_slice} show, the density of the tracer fluid correlates strongly with the gas density.

\subsubsection{Cooling Runs}\label{sec:results1b}

\begin{figure}
\begin{center}
\includegraphics[scale=0.4,trim=0cm 0cm 0cm 0cm, clip=true]{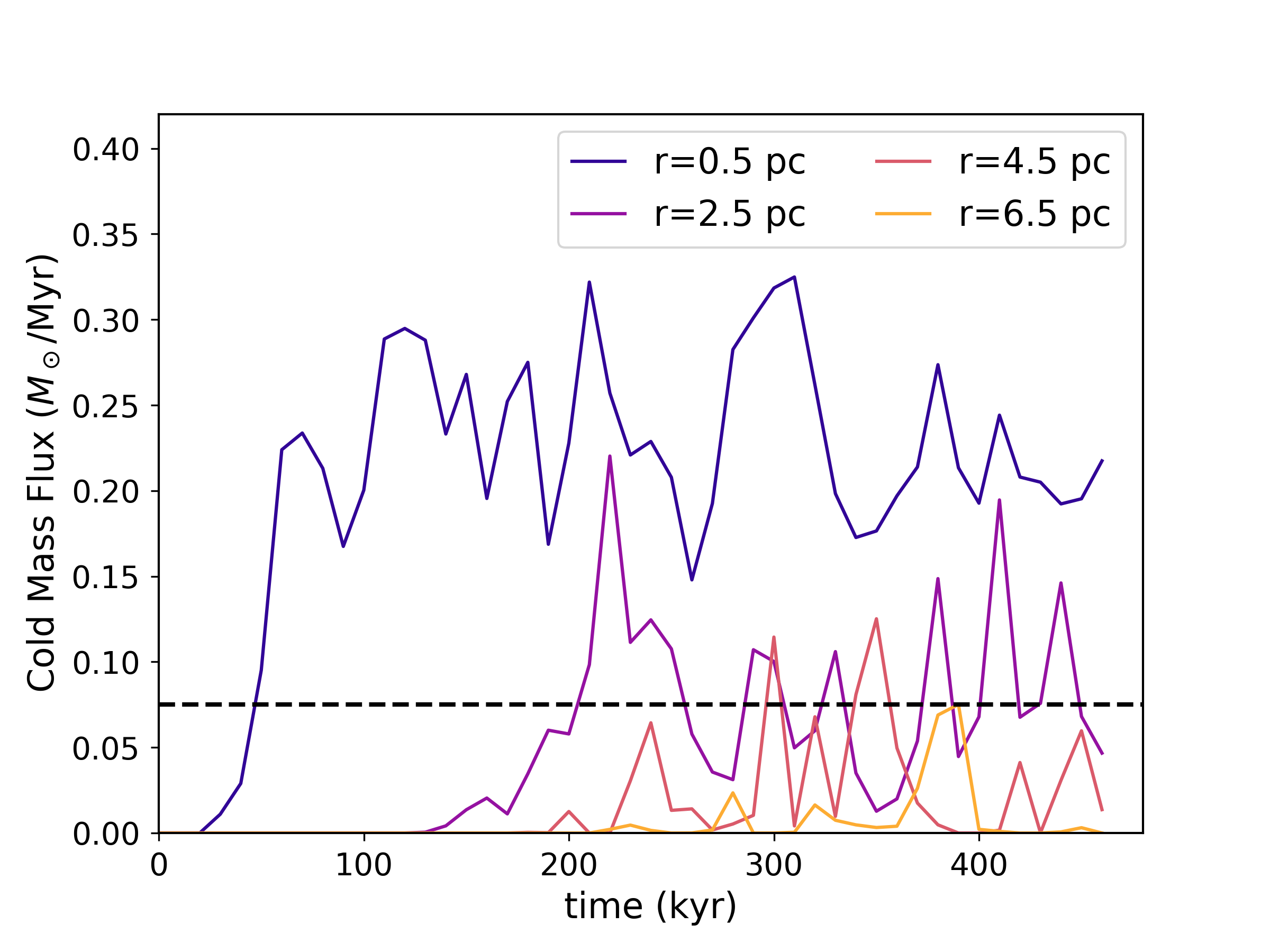}
\caption{Cold mass flux as a function of time at different distances from the star in the simulation of Mira's tail with radiative cooling.
\label{fig:Mira2_cold_flux}}
\end{center}
\end{figure}

\begin{figure}
\begin{center}
\includegraphics[scale=0.09,trim=2.1cm 0cm 1.5cm 0cm, clip=true]{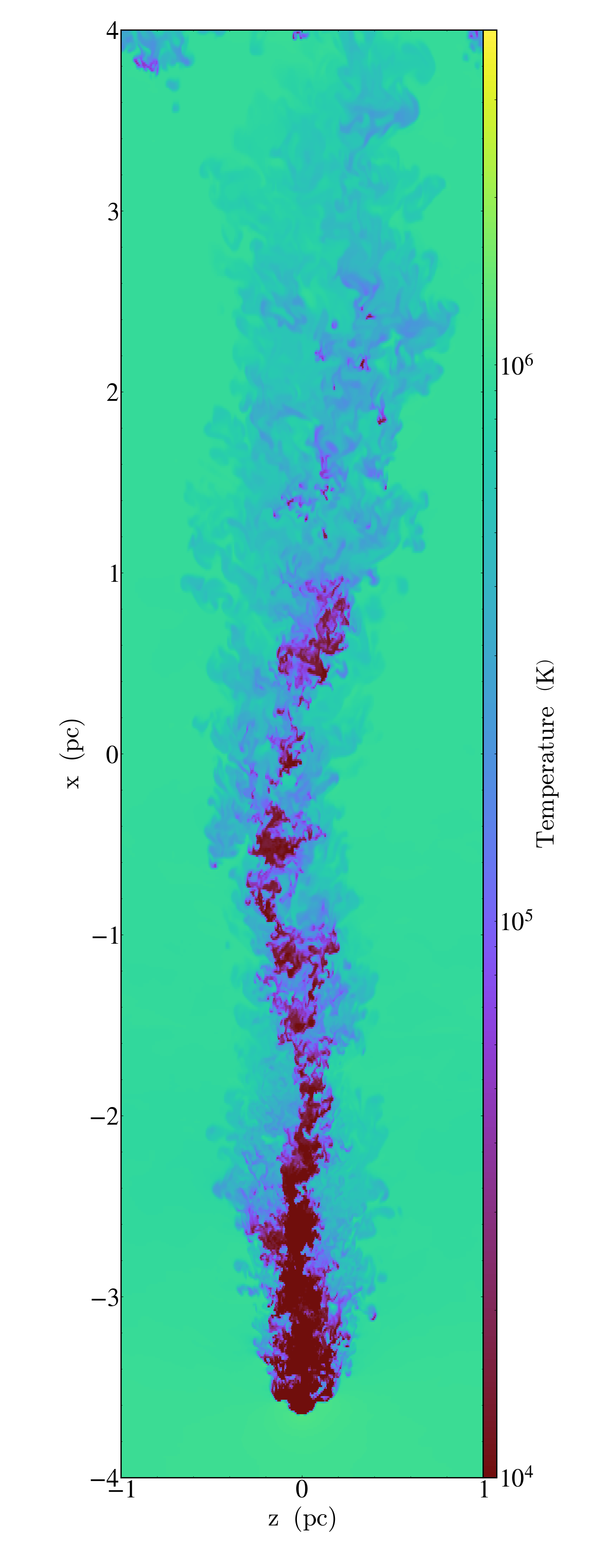}
\includegraphics[scale=0.09,trim=2.1cm 0cm 1.5cm 0cm, clip=true]{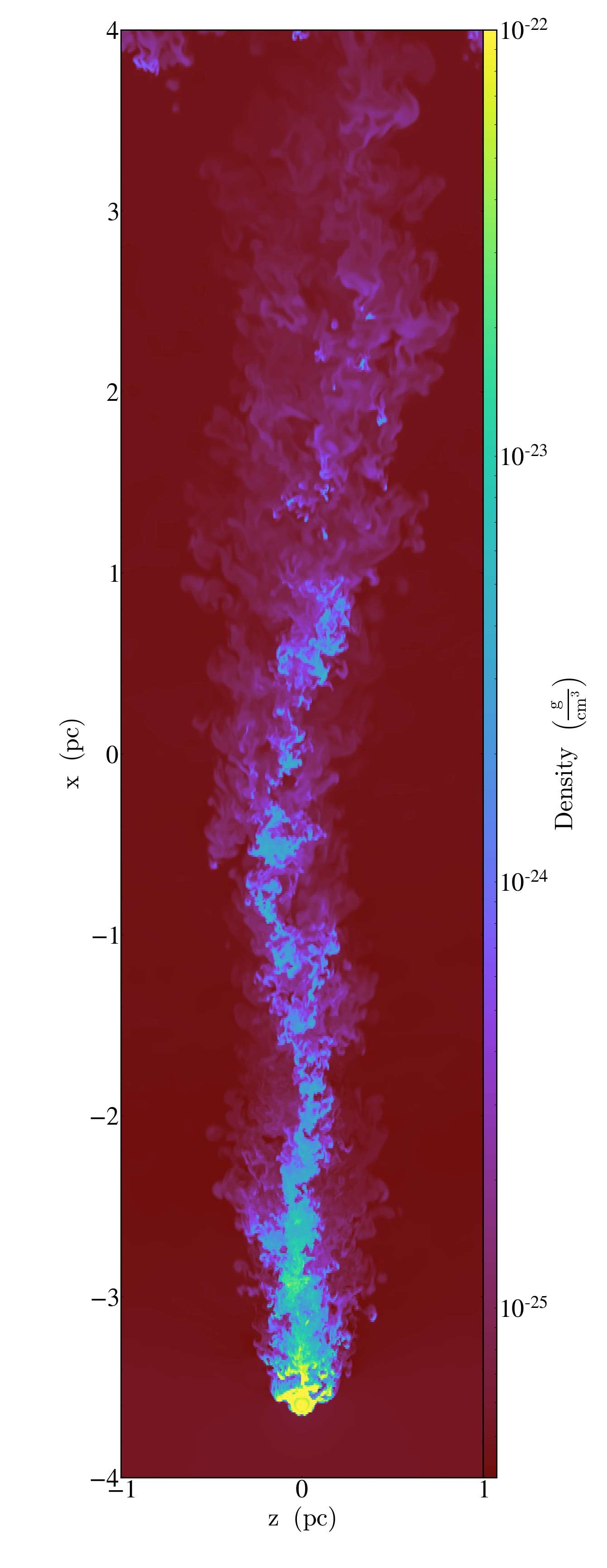}
\includegraphics[scale=0.09,trim=2.1cm 0cm 1.5cm 0cm, clip=true]{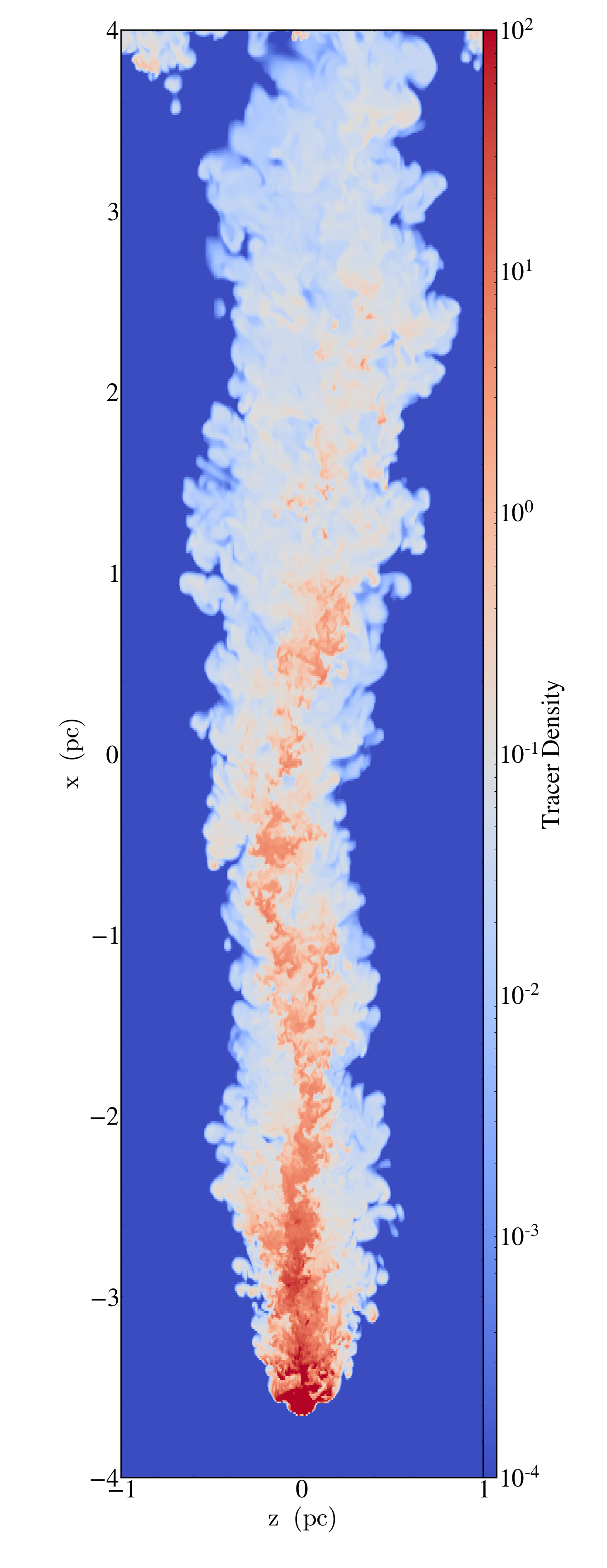}
\caption{Temperature, density and the density of tracer fluid in a slice of gas through the center of the simulation of Mira with cooling. 
\label{fig:Mira2_slice}}
\end{center}
\end{figure}

\begin{figure*}
\begin{center}

\includegraphics[scale=0.4,trim=0cm 0cm 0cm 1cm, clip=true]{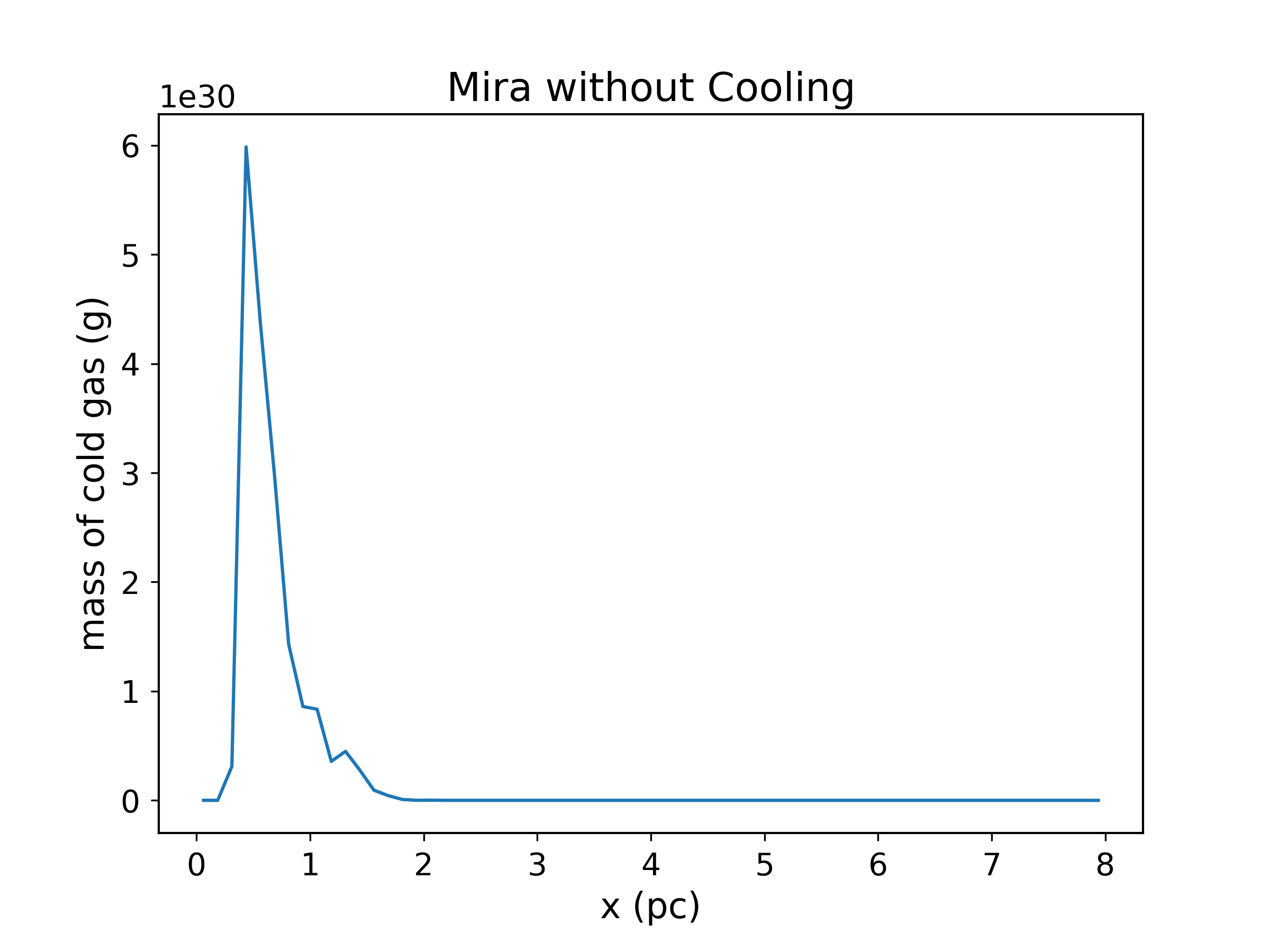}
\includegraphics[scale=0.4,trim=0cm 0cm 0cm 1cm, clip=true]{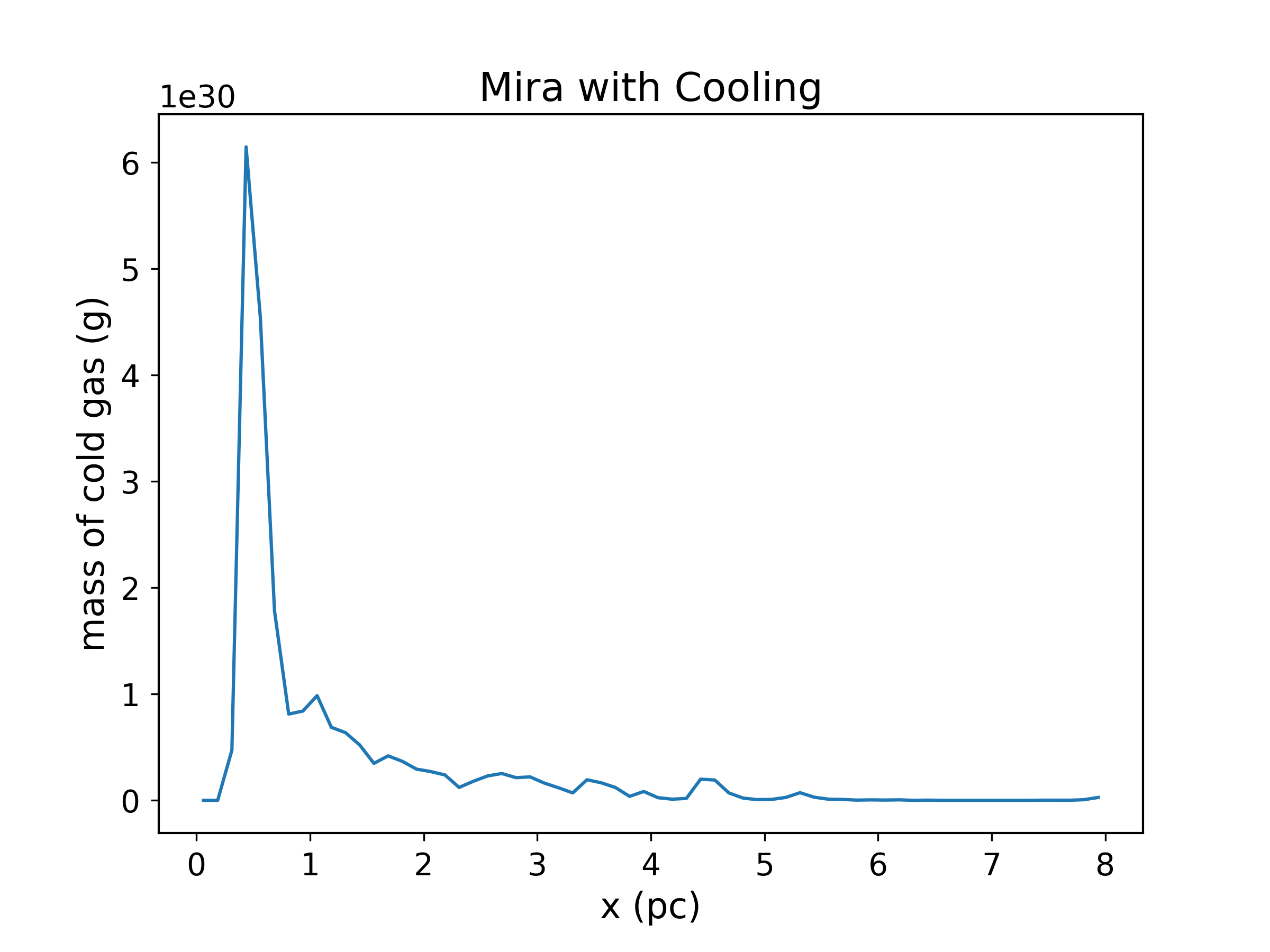}

\includegraphics[scale=0.4,trim=0cm 0cm 0cm 1cm, clip=true]{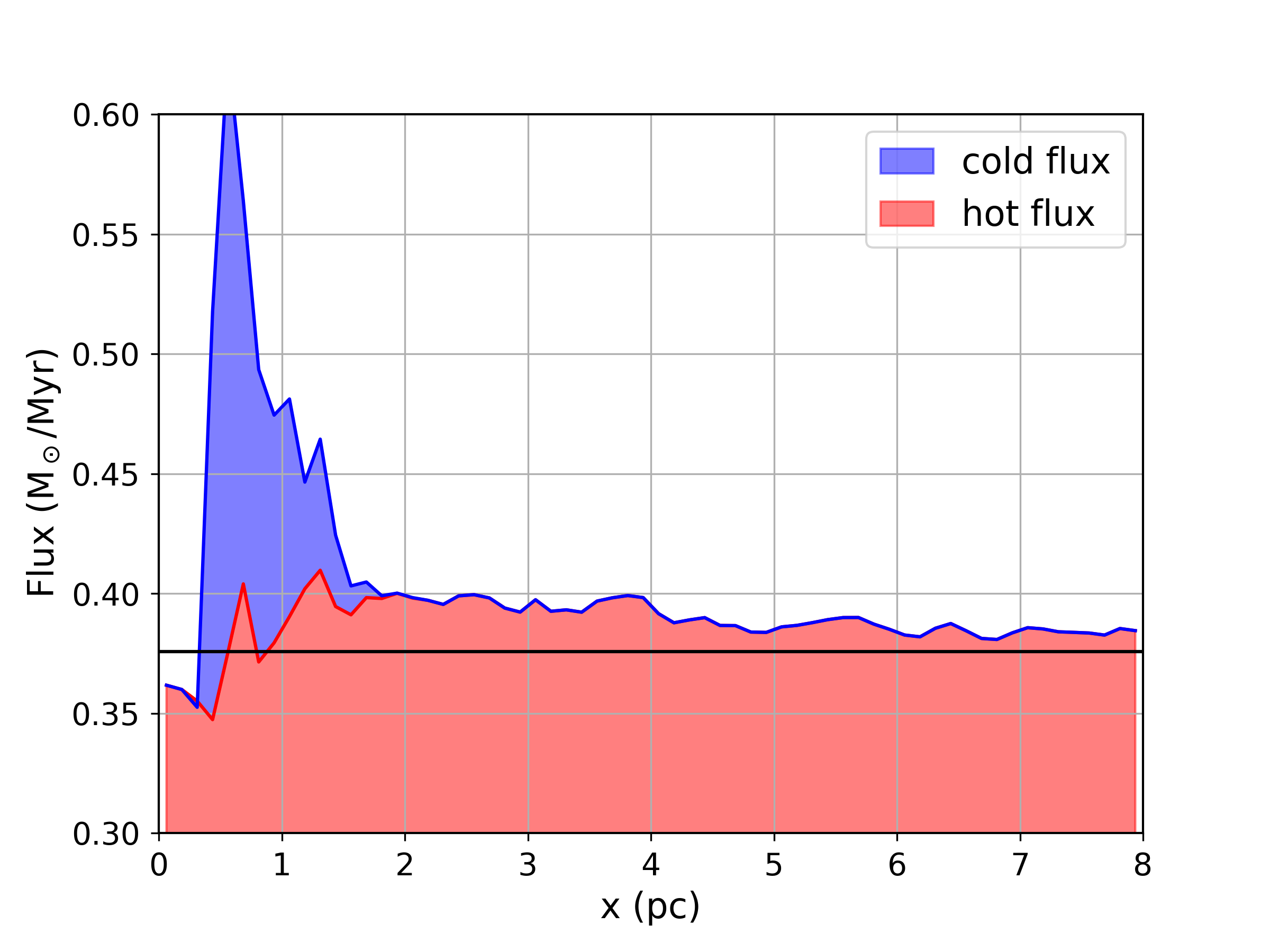}
\includegraphics[scale=0.4,trim=0cm 0cm 0cm 1cm, clip=true]{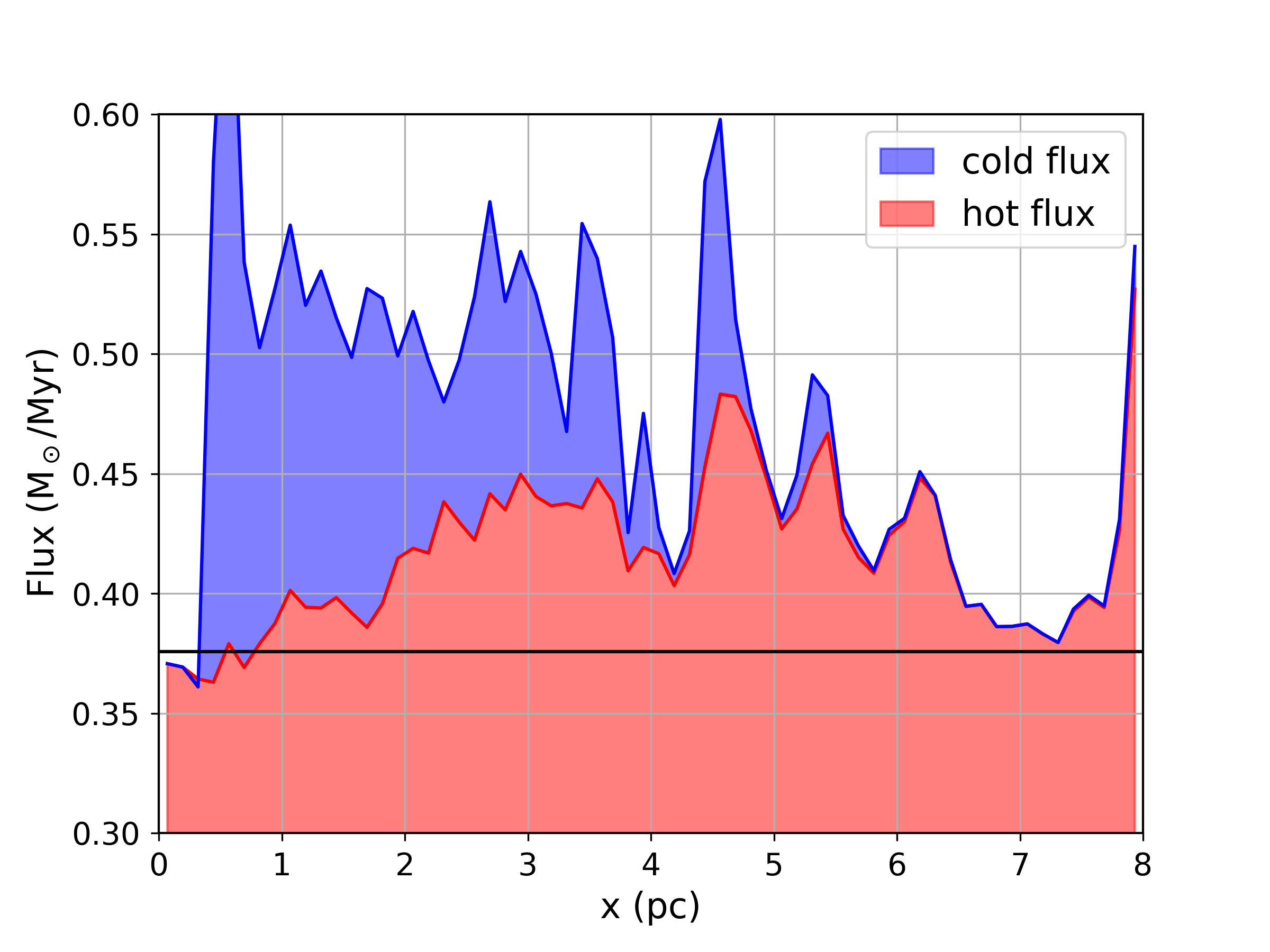}

\caption{Top panels: the mass of cold gas (defined as $T<2\times10^4 K$) as a function of x position in the simulation of Mira's tail without cooling (left) and with cooling (right). Bottom panels: the mass flux of cold and hot gas in the two simulations. The solid black line is the flux of the hot ISM that we inject into the left boundary of the simulation box. 
\label{fig:2by2}}
\end{center}
\end{figure*}

\begin{figure*}
\begin{center}
\includegraphics[scale=0.12,trim=0cm 0cm 0cm 0cm, clip=true]{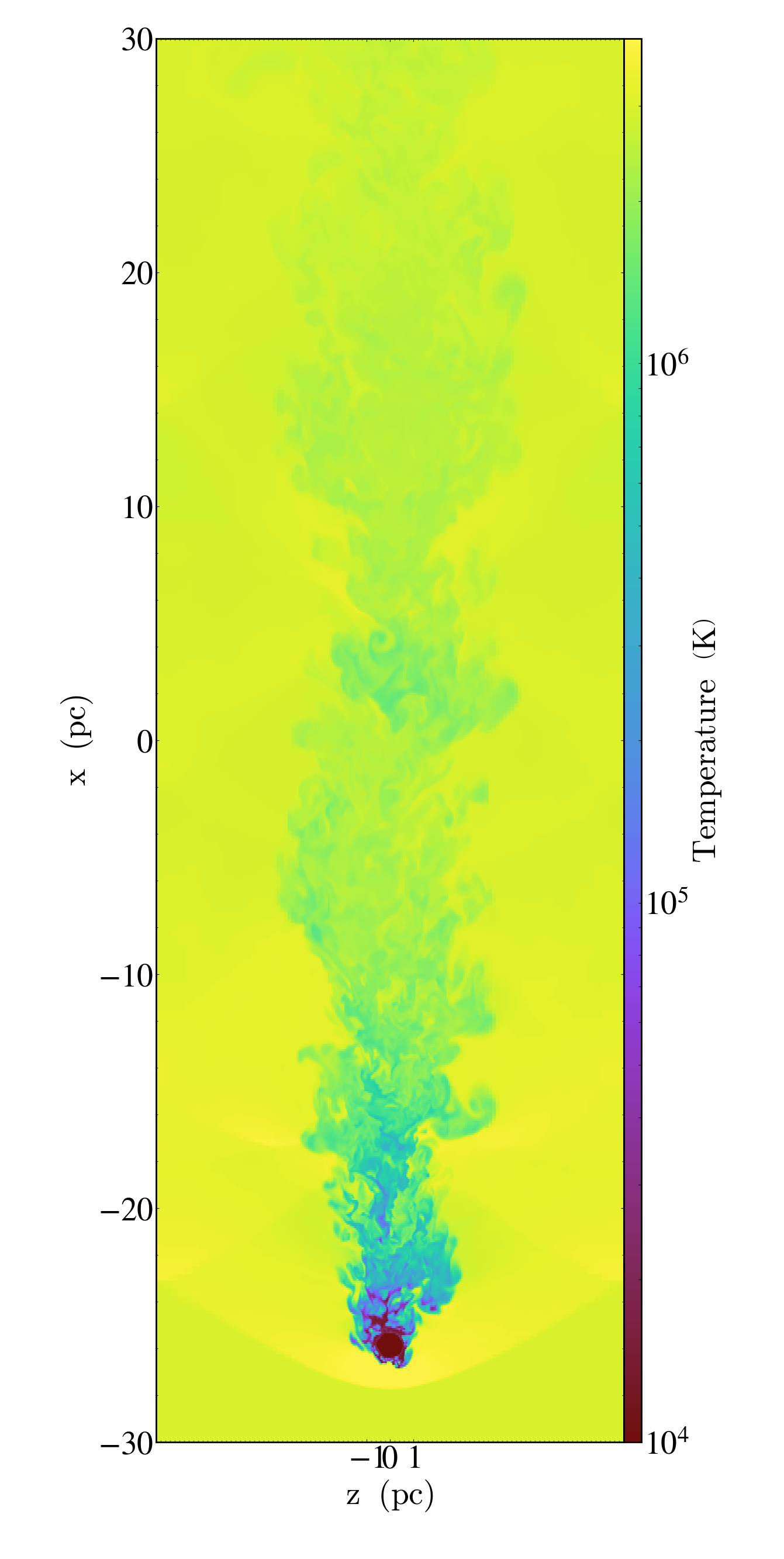}
\includegraphics[scale=0.12,trim=0cm 0cm 0cm 0cm, clip=true]{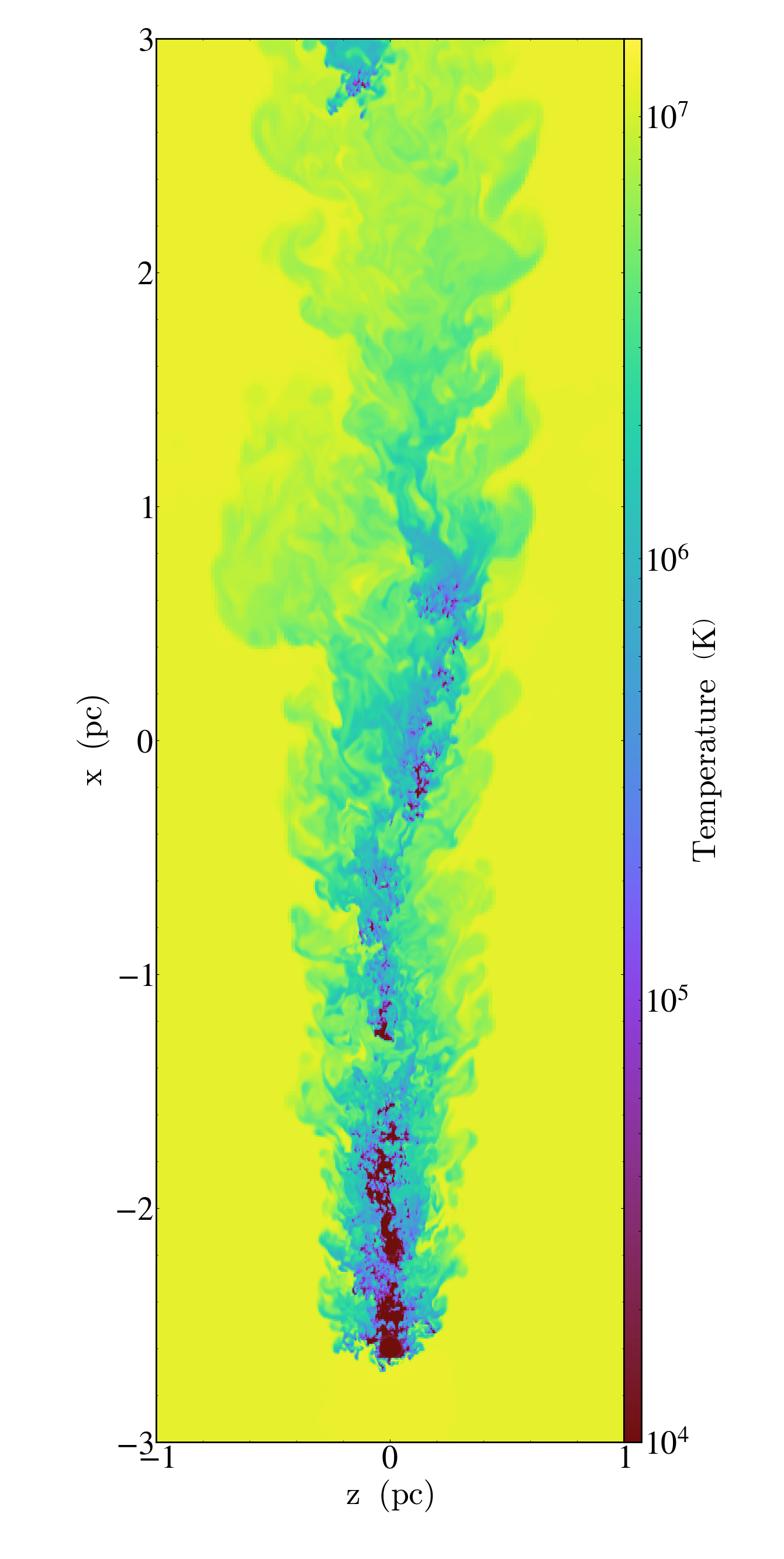}
\includegraphics[scale=0.12,trim=0cm 0cm 0cm 0cm, clip=true]{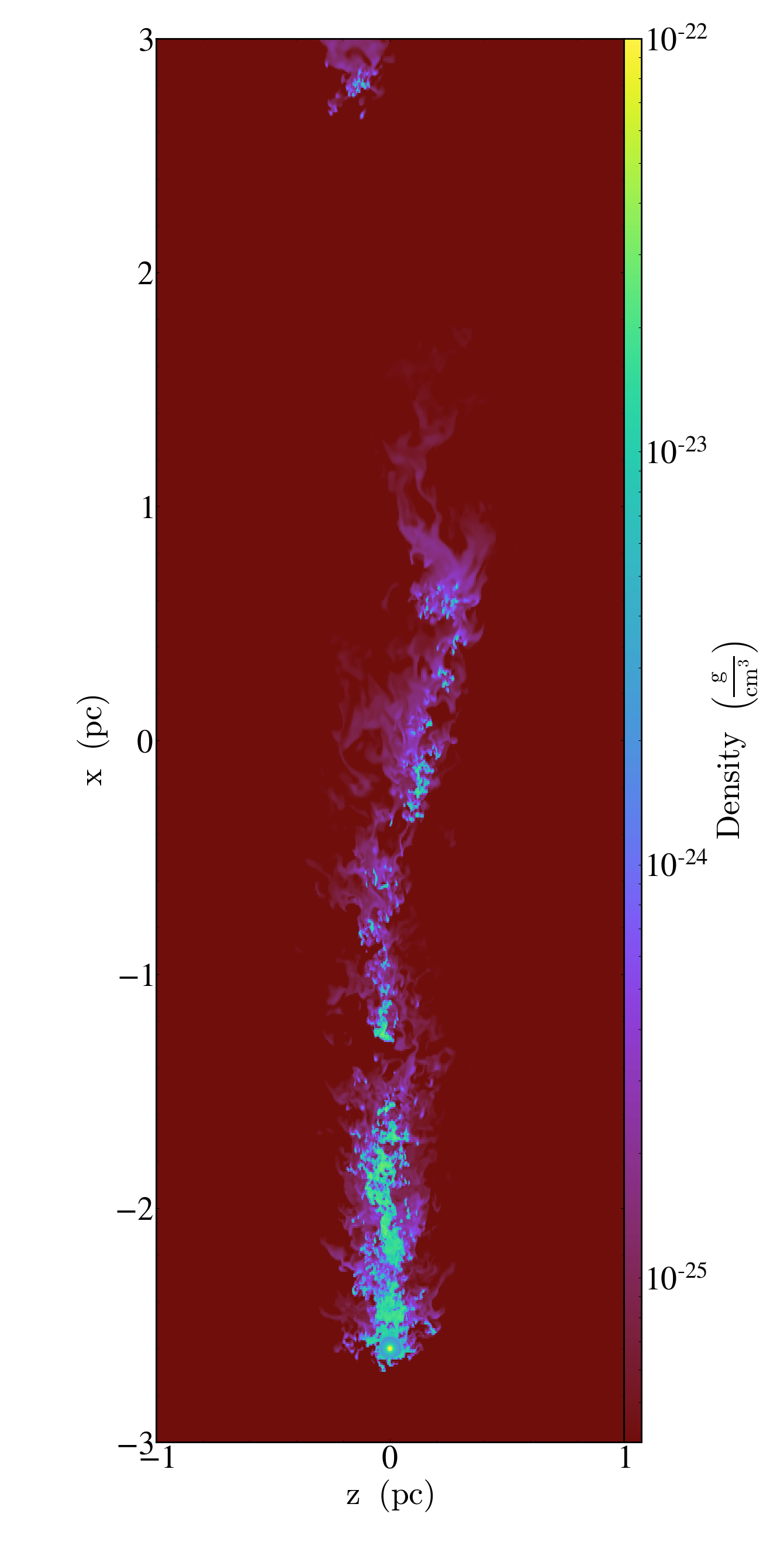}
\includegraphics[scale=0.12,trim=0cm 0cm 0cm 0cm, clip=true]{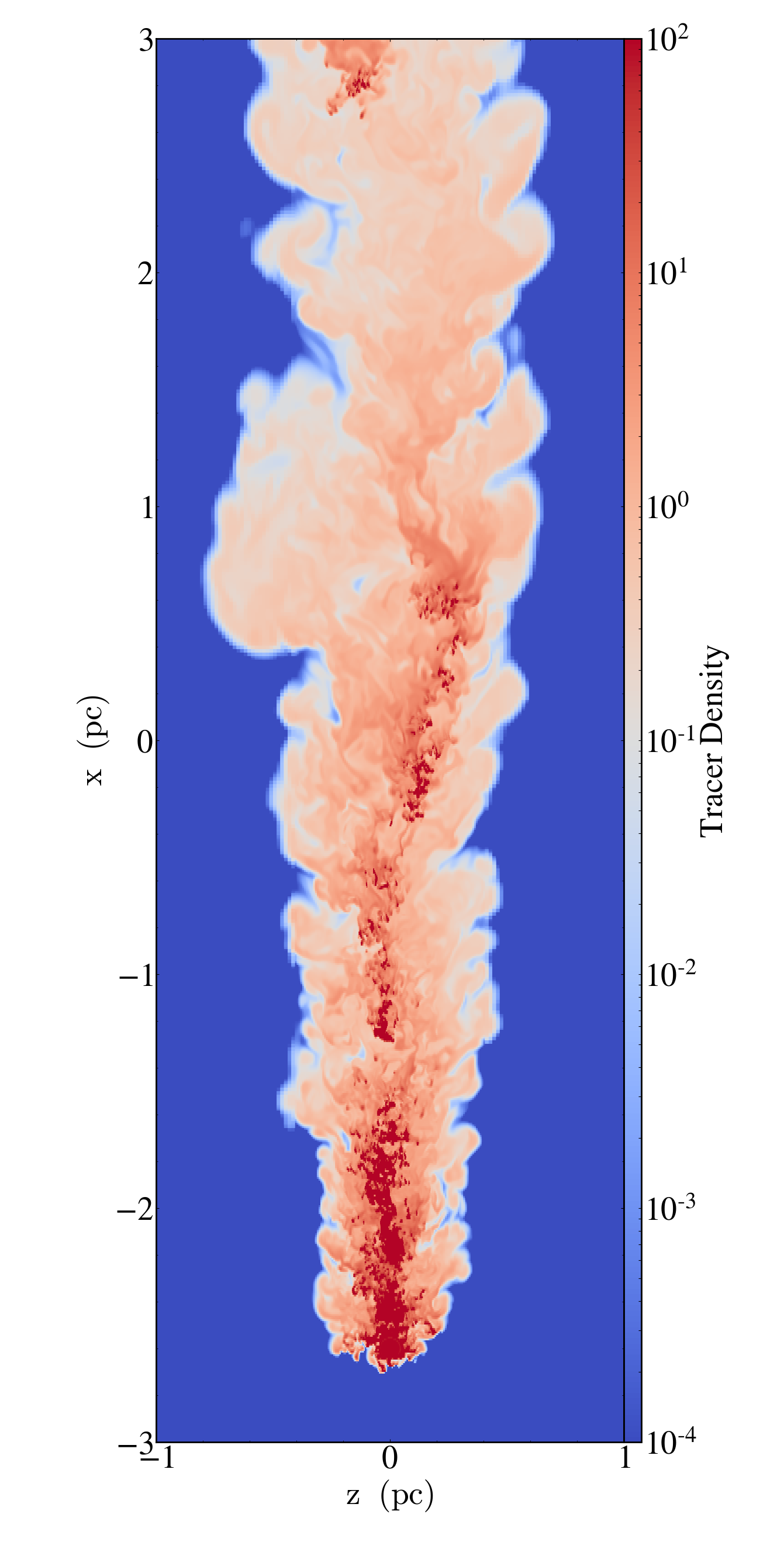}

\caption{Temperature slice of the LP run with cooling (left panel), and temperature, density and the density of tracer fluid in a slice of gas in the HP run with cooling (second to fourth panels). 
\label{fig:LPHP3_slice}}
\end{center}
\end{figure*}

In this section, we discuss the three simulations with radiative cooling. 

When cooling is allowed, the simulated cold tail of Mira becomes much longer (Figure~\ref{fig:Mira2_cold_flux}) and also slightly narrower (Figure~\ref{fig:Mira2_slice}), and is more consistent with the observed tail of Mira ($\sim 4$ pc long). As Figure~\ref{fig:2by2} shows, the amount of cold gas (defined as  $\rm T<2\times10^4 K$) drops to 0 before 2 pc in the adiabatic Mira simulation (left panels), whereas in the cooling run, the cold tail extends to $\sim 5$ pc (right panels). In both simulations, the flux of hot gas ($\rm T>2\times10^4 K$) increases as the cold AGB wind gets mixed with the hot ISM. In the simulation with cooling, the amount of cold gas in the tail decreases with distance from the star, but at a smaller rate than the adiabatic run. Occasionally, small clumps of cold gas are even seen leaving the simulation box (e.g., Figure~\ref{fig:Mira2_slice}). 

However, this is not the case for the low pressure run. In the low pressure run, the properties of the tail are essentially identical to the adiabatic run. For the high pressure run though, the tail becomes long again, and appears very similar to Mira's tail. 

This is consistent with our finding in Section~\ref{sec:analytic} that the cooling time of the tail is inversely proportional to the ambient pressure. Mira's ambient environment has a similar pressure to the high pressure run (the temperature of the local bubble is lower but the density is higher), and thus cooling is important. The low pressure run has an ambient pressure more than an order of magnitude smaller than the other two cases. Therefore radiative cooling is unimportant and does not affect the properties of the tail.

\subsection{Tail Properties (Mira)}\label{sec:tail}
In this section, we examine the mixing and cooling processes in the tail of Mira in more detail, focusing on the difference between the adiabatic and cooling runs. Although the figures and analysis are for Mira, the HP run is very similar.

\begin{figure*}
\begin{center}

\includegraphics[scale=0.4,trim=0cm 0cm 0cm 1cm, clip=true]{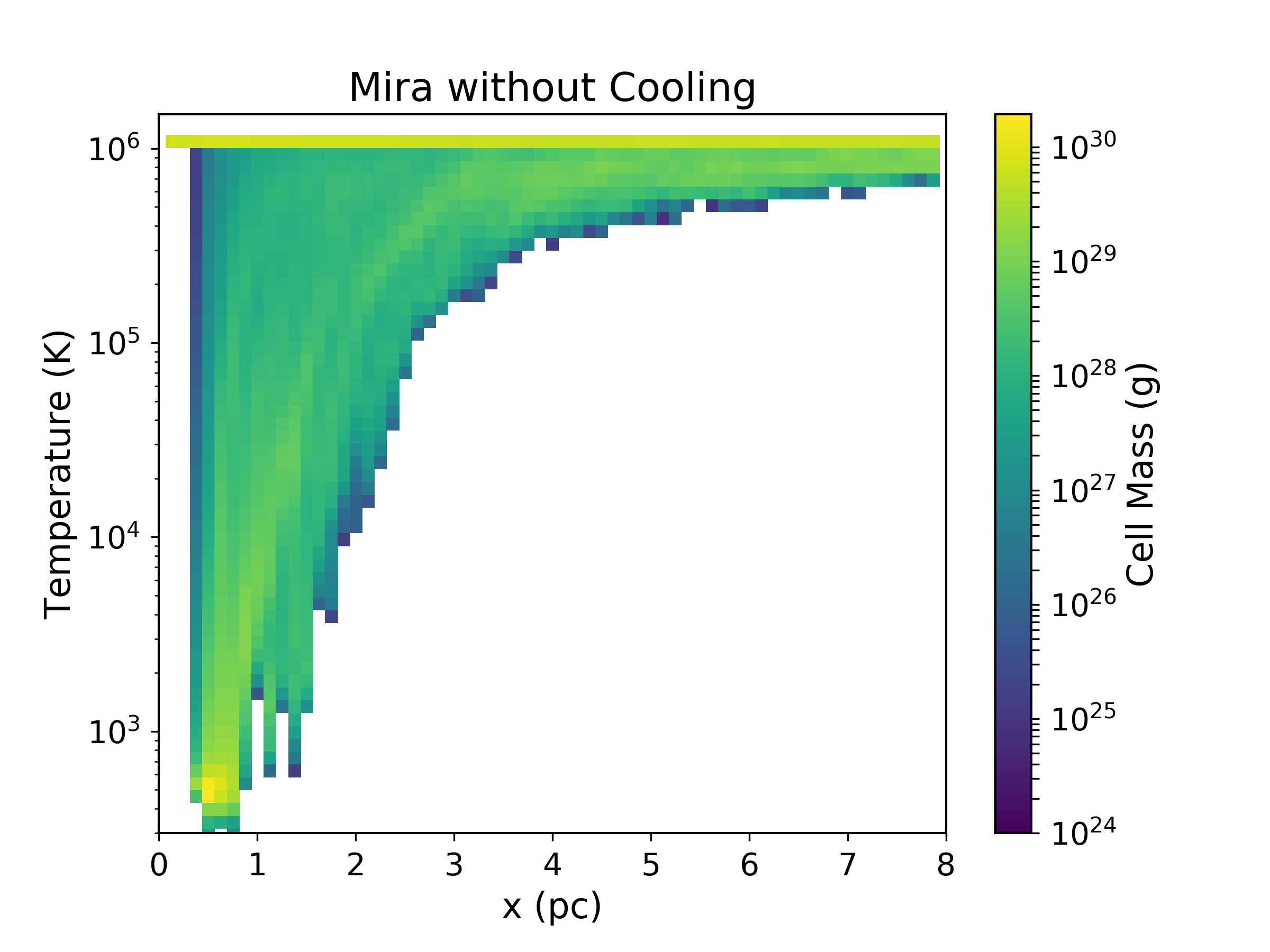}
\includegraphics[scale=0.4,trim=0cm 0cm 0cm 1cm, clip=true]{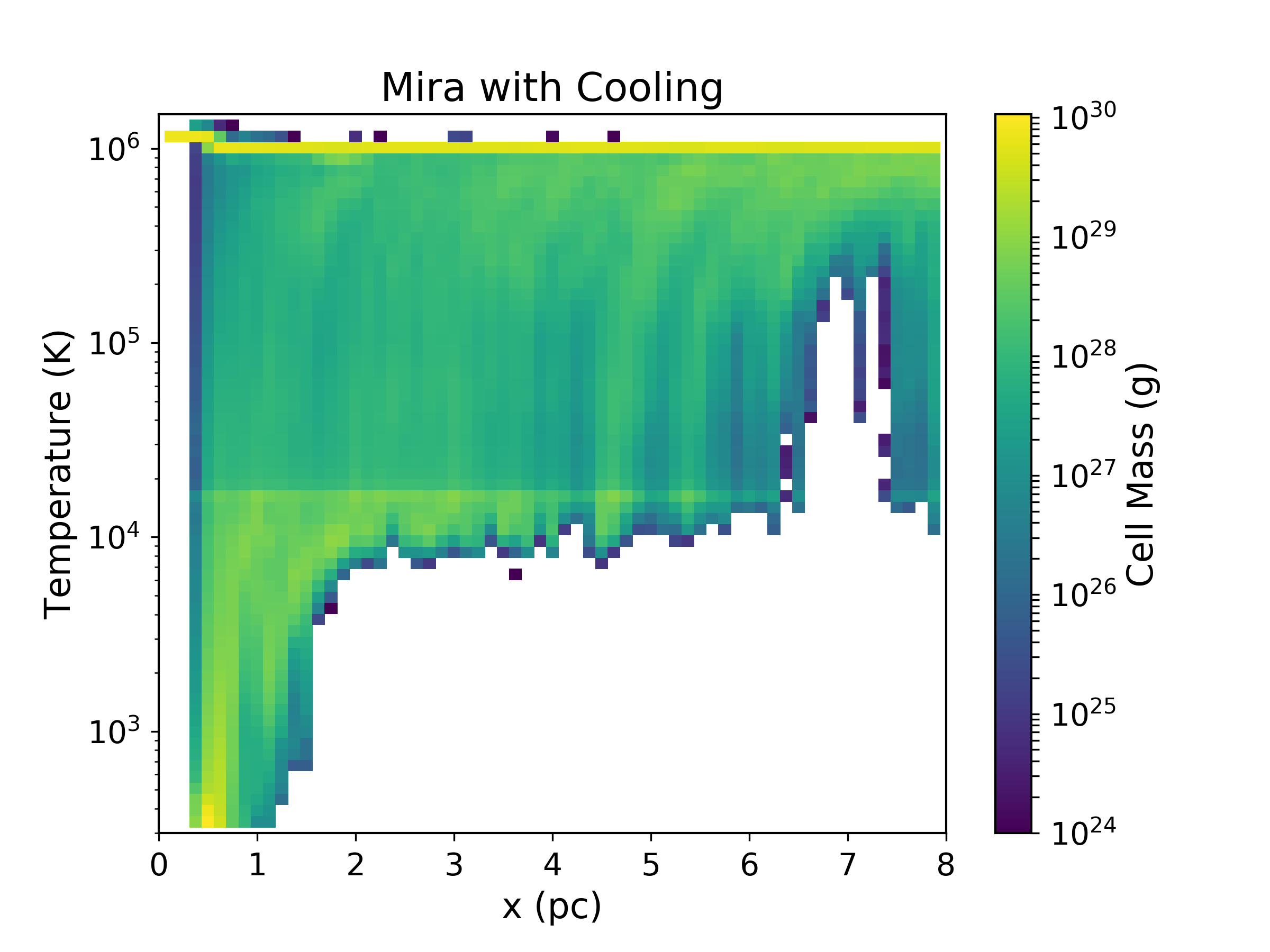}

\includegraphics[scale=0.4,trim=0cm 0cm 0cm 1cm, clip=true]{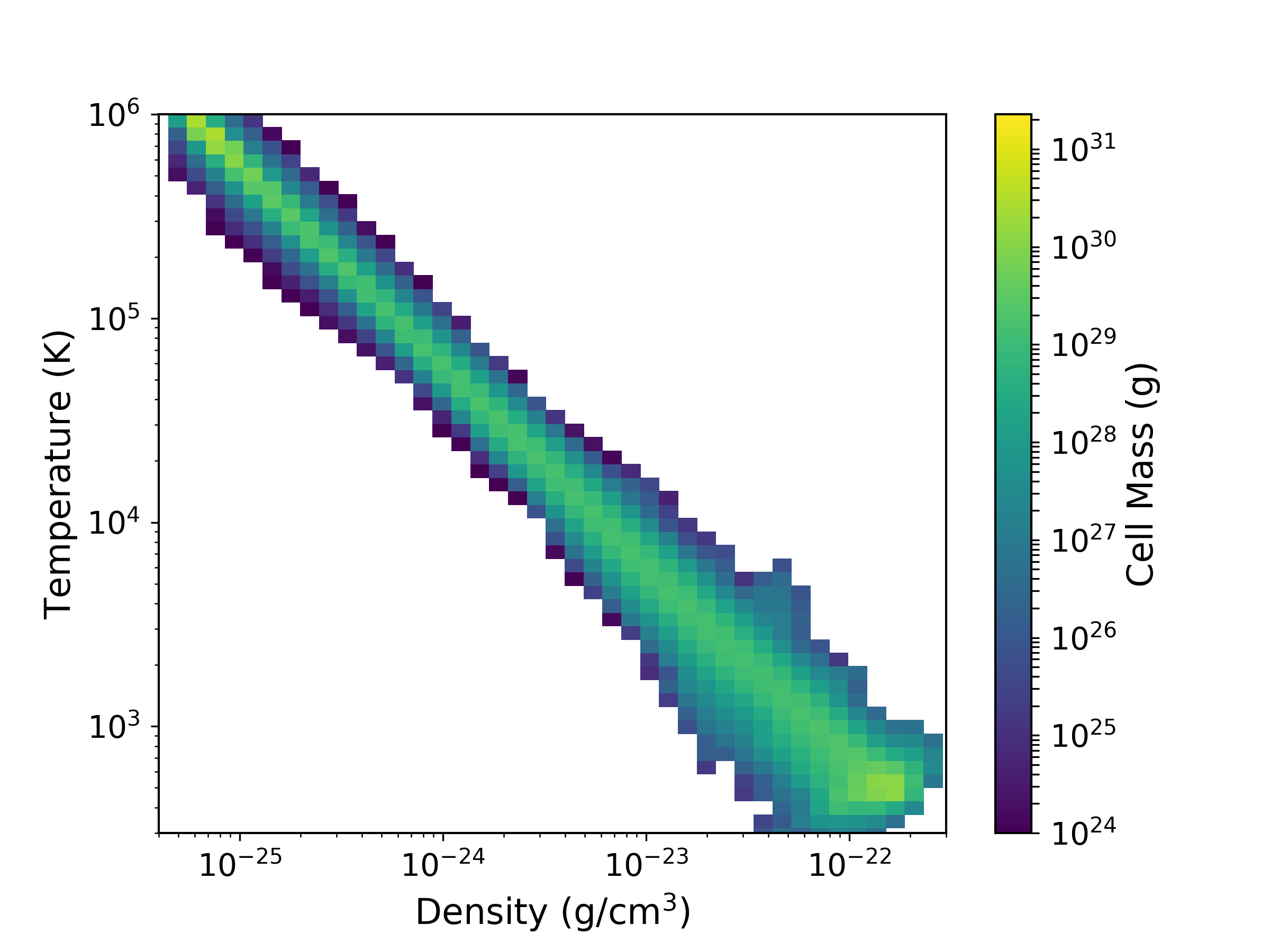}
\includegraphics[scale=0.4,trim=0cm 0cm 0cm 1cm, clip=true]{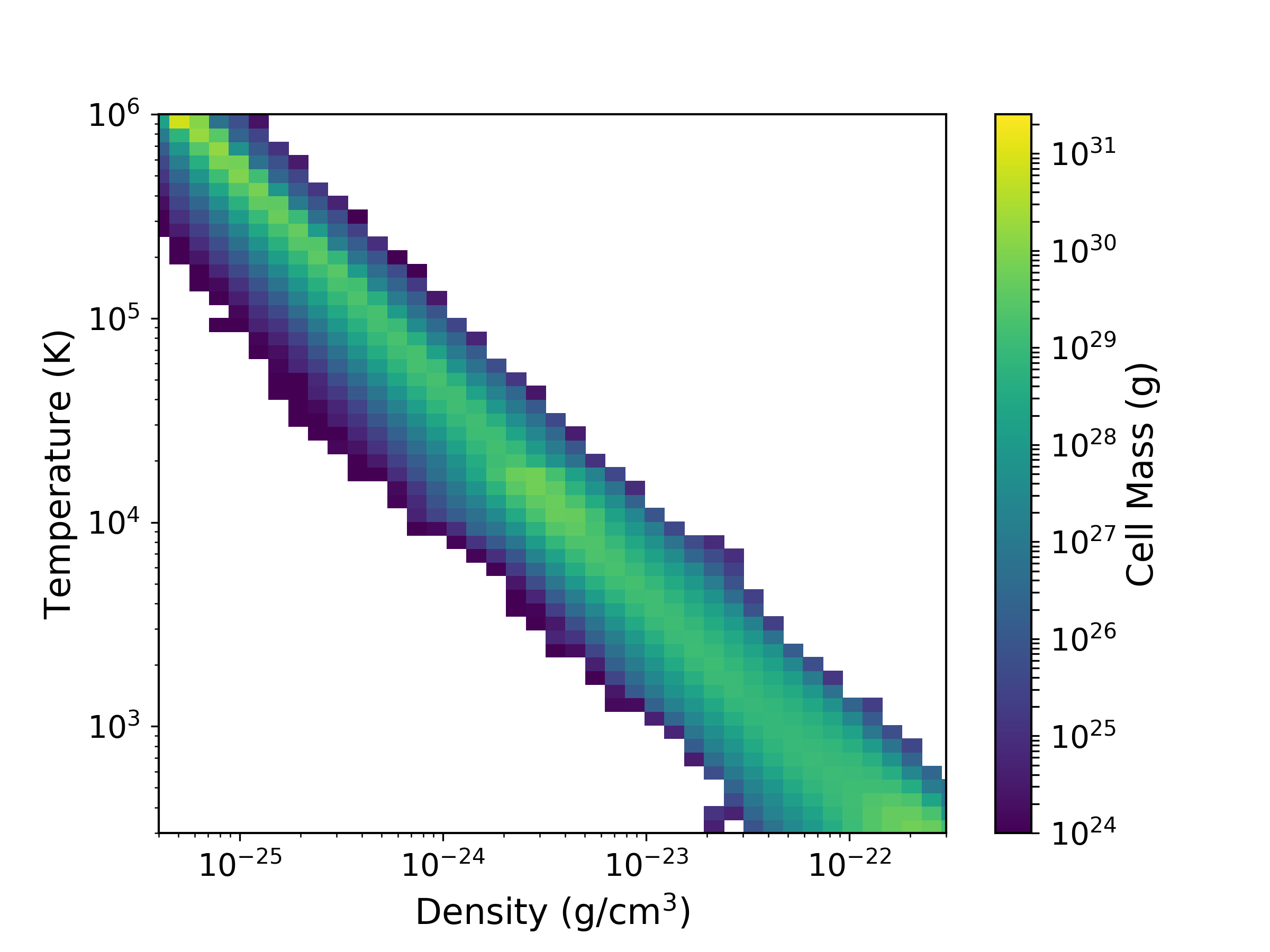}

\includegraphics[scale=0.4,trim=0cm 0cm 0cm 1cm, clip=true]{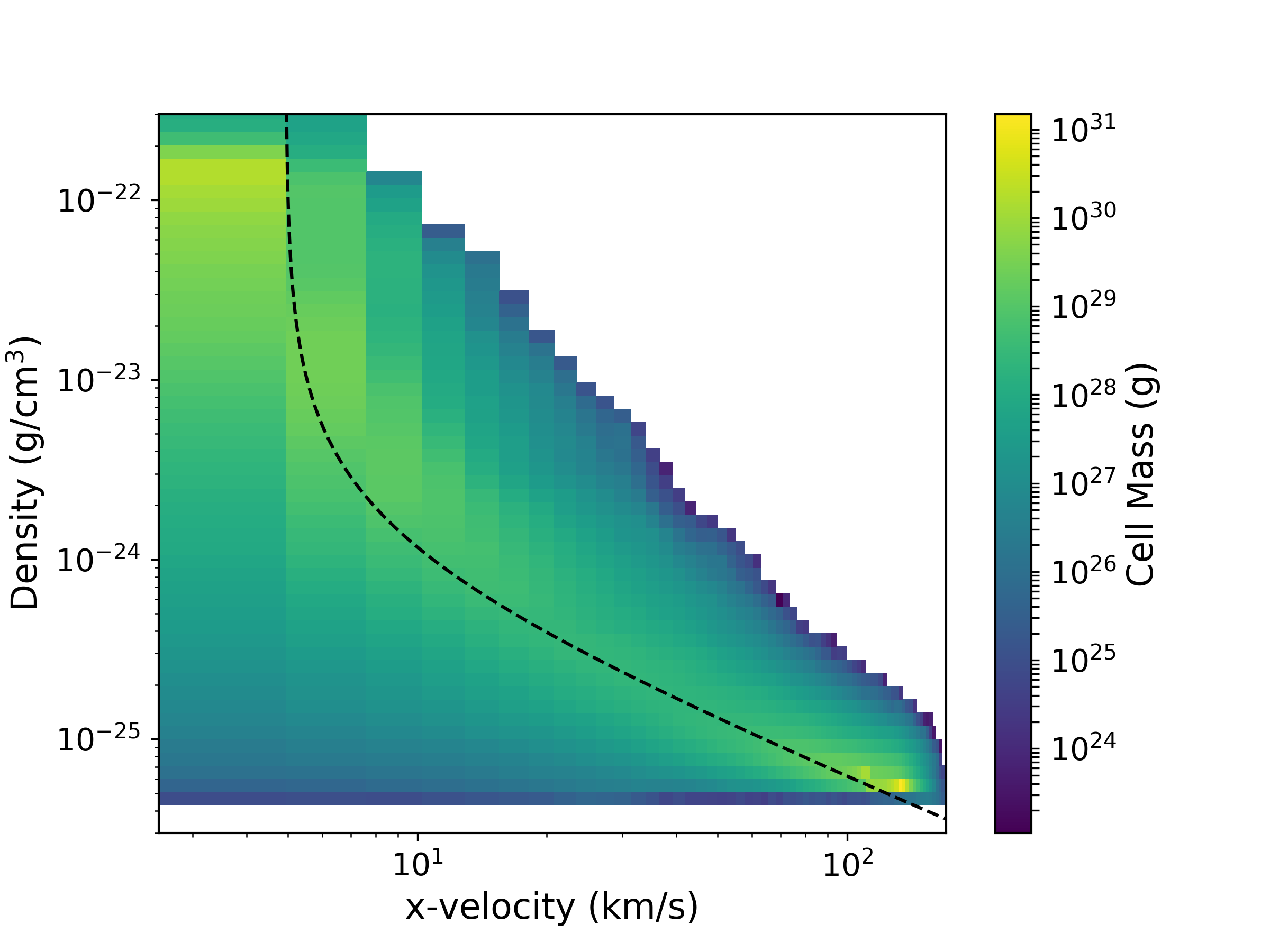}
\includegraphics[scale=0.4,trim=0cm 0cm 0cm 1cm, clip=true]{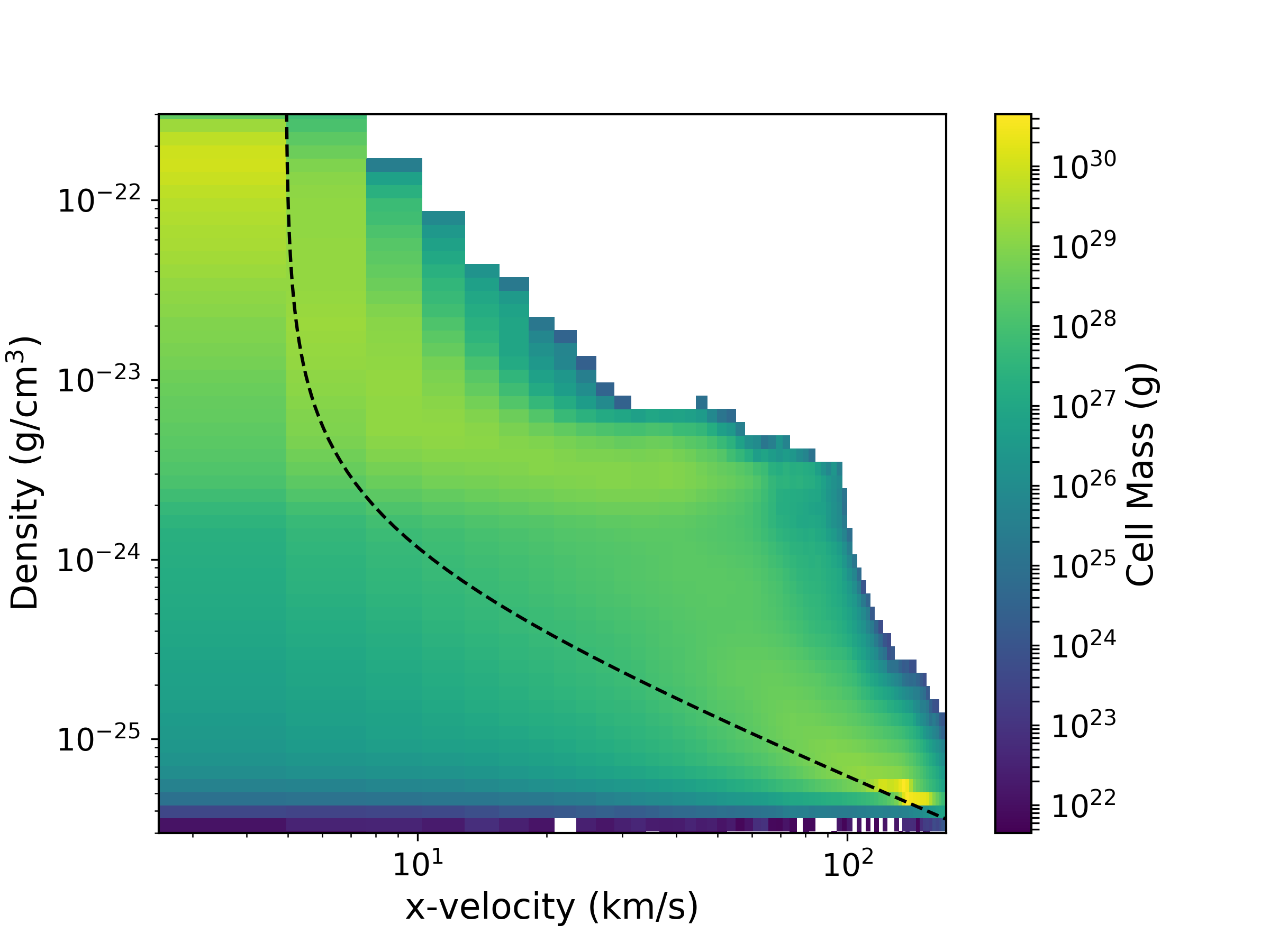}

\includegraphics[scale=0.4,trim=0cm 0cm 0cm 1cm, clip=true]{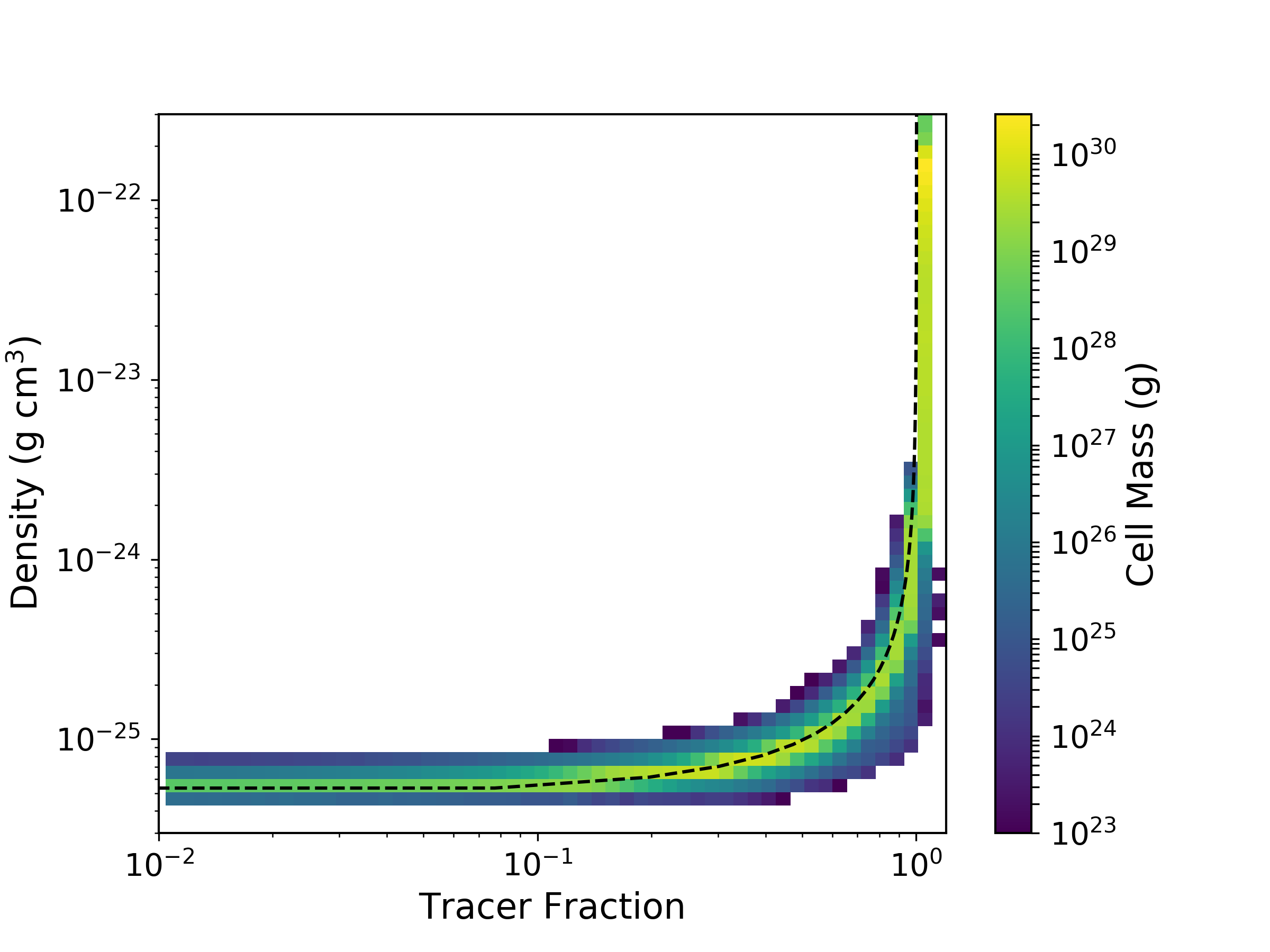}
\includegraphics[scale=0.4,trim=0cm 0cm 0cm 1cm, clip=true]{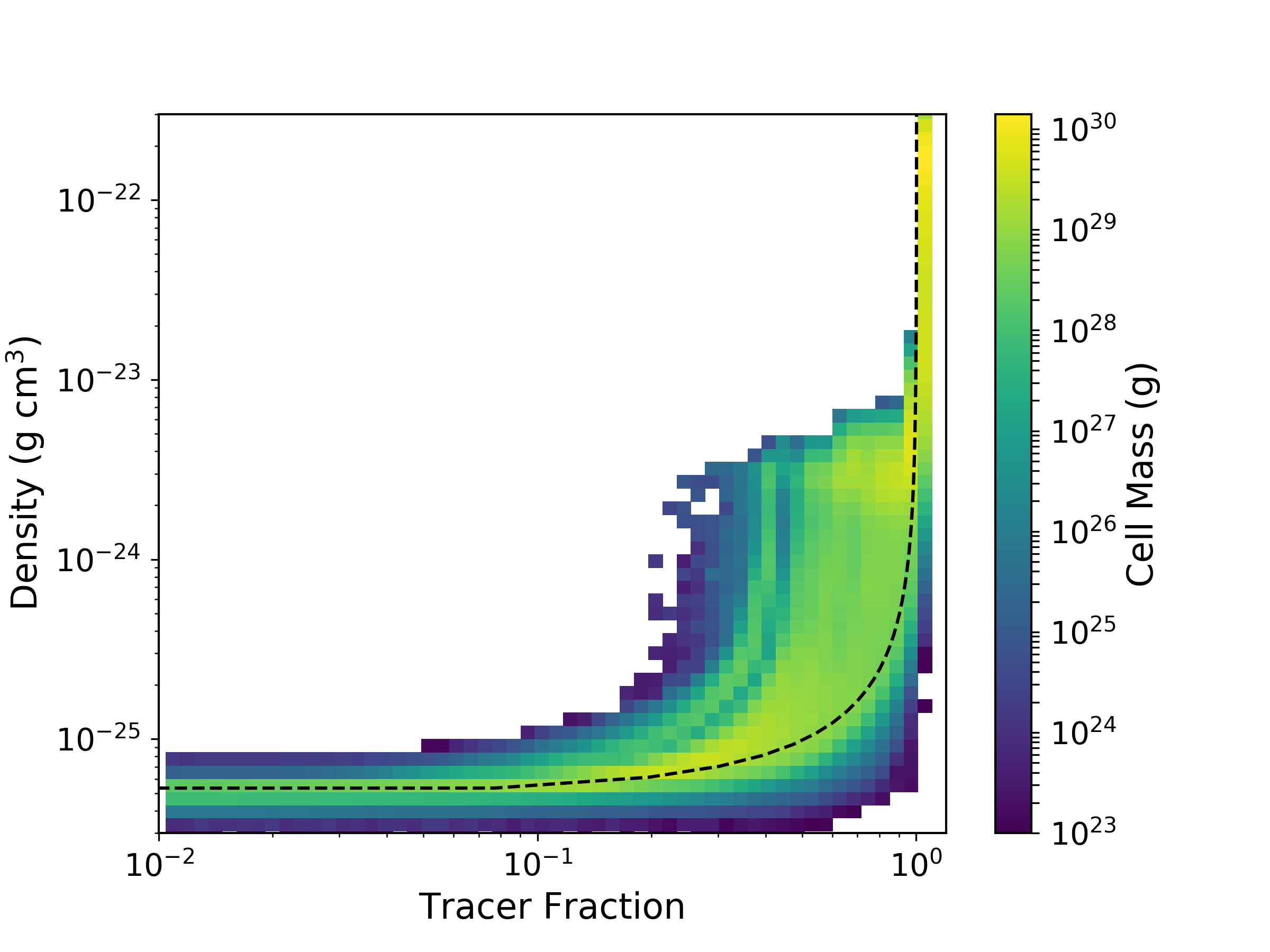}

\caption{Top panels: temperature of gas as a function of x position. Second row: gas density-temperature distribution. Third row: gas density as a function of x-velocity. Bottom panels: gas density as a function of tracer fraction. All panels on the left are for simulations of Mira without radiative cooling, and panels on the right are with cooling. Dashed lines are analytical calculations for pure mixing (see Section~\ref{sec:tail} for detail). 
\label{fig:phase}}
\end{center}
\end{figure*}

Figure~\ref{fig:phase} shows (from top to bottom) the distribution of gas temperature as a function of x-coordinate (along the motion of the hot ISM wind), density-temperature diagram, velocity-density diagram, and the distribution of gas density as a function of the tracer fraction. In the adiabatic run, the temperature distribution becomes narrower and narrower as the cold AGB wind is mixed into the hot ISM downstream (top left panel). In the cooling run (top right panel), the temperature of the mixing AGB wind reaches $\sim 10^4$ K just as quickly from its original temperature of a few hundred K. Then a significant fraction of the gas stays at $\sim 10^4$ K due to cooling of the gas with higher temperatures and high cooling rate (at $\sim 10^5$ K). The gas does not cool further because at $10^4$ K the cooling rate becomes low. The amount of $10^4$ K gas decreases with distance, but not smoothly, as the tail is rather turbulent. In this snapshot, some cold gas is seen leaving the simulation box (see also Figure~\ref{fig:Mira2_slice}). 

The second row of Figure~\ref{fig:phase} shows the density-temperature distribution of the gas in the two simulations. In both cases, temperature and density are inversely proportional to each other, suggesting that the gas is mostly in pressure equilibrium. In the cooling run (right panel), the distribution is slightly broader, especially between $10^4-10^5$ K where the cooling rate peaks, indicating that some of the rapidly cooling gas is slightly under-pressured. Recent numerical simulations focusing on the mixing layers have also found that the cooling mixing layer has a slightly lower thermal pressure \citep{Ji2018}. Overall, the pressure variation is small, and we will assume that the gas is in pressure equilibrium in the following analysis.

If mixing is the only important physical process, we can estimate the density and velocity of the mixed material ($\rho_{mix}$ and $v_{mix}$) as a function of the normalized tracer density $\chi$, defined as the density of the tracer fluid $\rho_c$ divided by its density at the standoff radius $\rho_0$. In the simulations, we set the tracer density at launching to be equal to the gas density, and thus $\rho_0=\rho_w$. During isobaric mixing, in the absence of cooling, both mass and internal energy are conserved, and we have:
\begin{equation}
\rho_{mix}=\chi \rho_w+(1-\chi)\rho_{ISM}\, .
\end{equation}

Momentum conservations gives us:
\begin{equation}
\rho_{mix}v_{mix}=\chi \rho_w v_w+(1-\chi)\rho_{ISM}v_{ISM}\, .
\end{equation}

Thus the velocity of the mixed gas as a function of its density is 
\begin{equation}\label{eq:rhov}
v_{mix}=\frac{\rho_wv_w-\rho_{ISM}v_{ISM}}{\rho_w-\rho_{ISM}} \Big(1-\frac{\rho_{ISM}}{\rho_{mix}}\Big) +\rho_{ISM}v_{ISM} \,.
\end{equation}

The third panels of Figure~\ref{fig:phase} show the distribution of all the gas in the simulation box on the velocity-density phase diagram in the adiabatic run (left) and the cooling run (right). The black dashed line is the analytical expectation from pure mixing (Equation~\ref{eq:rhov}). It agrees well with the adiabatic run, but in the run with cooling, there is an excess of material with intermediate velocity but higher density as a result of condensation of the mixed gas. 

The effect of cooling can be also seen in the bottom panels of Figure~\ref{fig:phase}, where we plot the phase diagram of tracer fraction-density. Tracer fraction $f$ is defined as the density of the tracer fluid divided by the density of the gas: $f=\rho_c/\rho_{mix}$. Its relation to $\rho_{mix}$ can be expressed as
\begin{equation}
f=\frac{\rho_w}{\rho_w-\rho_{ISM}}\Big(1-\frac{\rho_{ISM}}{\rho_{mix}} \Big)
\end{equation}
This analytical expectation, plotted as a black dashed line, again agrees with the adiabatic simulation well. There is a tight correlation between the density of the mixed gas and the tracer fraction. In the cooling run, some of the gas with intermediate tracer fraction (ISM that is partially polluted by the AGB wind) has condensed. Because cooling rate is proportional to $n^2$, the most diffuse ISM that is least contaminated by the AGB wind (with low tracer fraction) does not cool and follows the adiabatic mixing line.

\section{Discussions}
\label{sec:discussions}

\subsection{Comparison with Observations}\label{sec:obs}
In this section, we compare our simulated Mira with the observations. As stated earlier, the focus of this work is not precisely reproducing Mira. Because we do not follow the creation and destruction of molecules, we cannot predict the exact emissivity. We simply assume that the density of the cold gas in the simulation is linearly proportional to the density of H$_2$ molecules which presumably produce the UV photos detected by GALEX \citep{Martin2007}. We will leave a more comprehensive comparison with Mira observations for future work when we include a more realistic model for the time evolution of AGB mass loss rate and a more sophisticated model for chemical evolution. Here we focus on two aspects: the velocity of Mira's tail and its periodicity. 

\begin{figure}
\begin{center}
\includegraphics[scale=0.4,trim=0cm 0cm 0cm 0cm, clip=true]{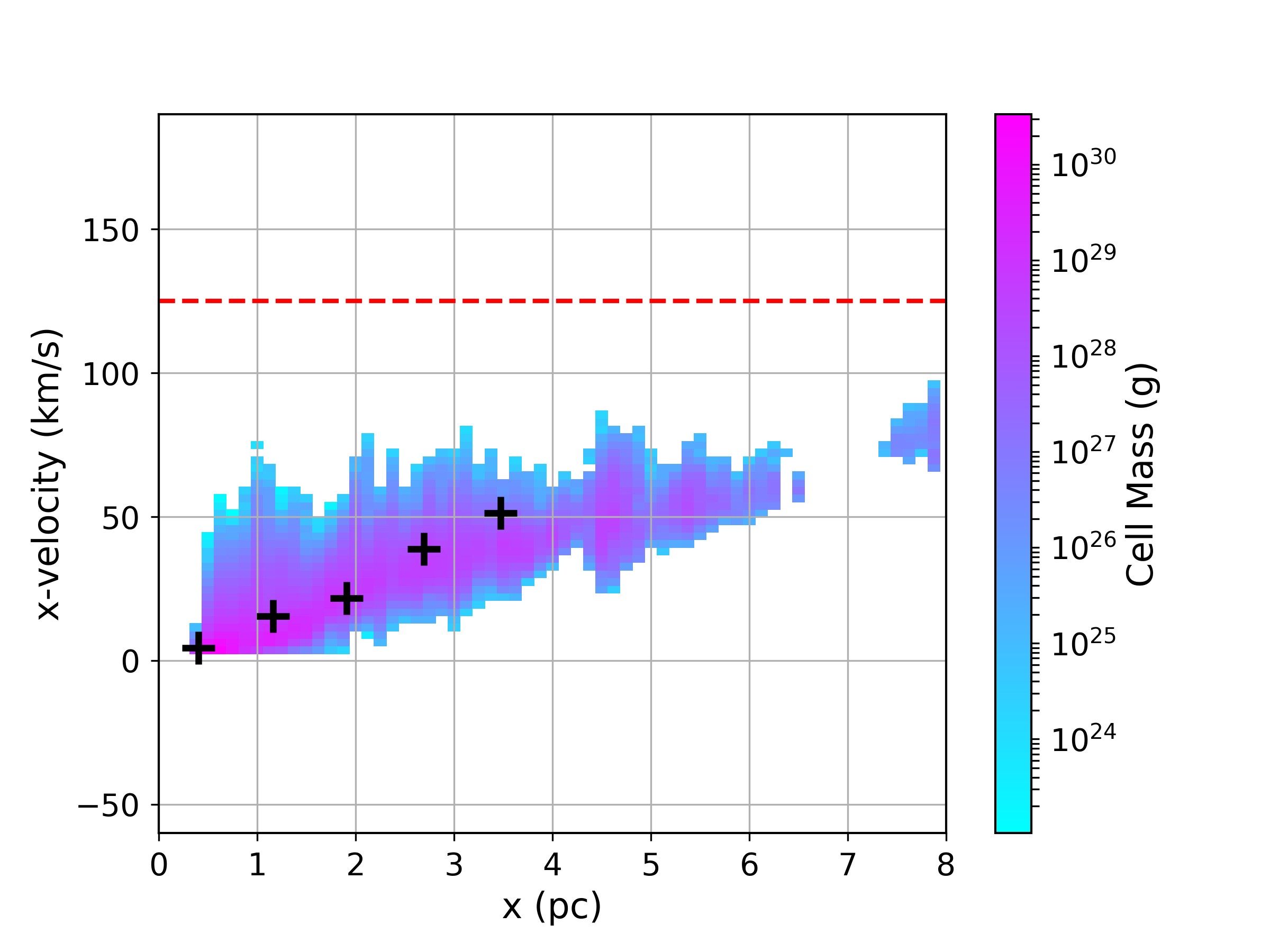}
\caption{The velocity of the cold gas (defined as $T<2\times10^4$ K) as a function of x position in the Mira simulation with cooling. Black crosses are data from HI observations of Mira's tail \citep{Mathews2008}. Dashed line is the velocity of the hot ISM. 
\label{fig:HI}}
\end{center}
\end{figure}

\citet{Mathews2008} measure the velocity of Mira's tail using HI observations. In Figure~\ref{fig:HI}, we show the velocity-position phase diagram of our Mira simulation with cooling. The observational data from \citet{Mathews2008} are over-plotted as black symbols, and agree with our simulation well. \citet{Raga2008} use a ``turbulent wake'' model to explain the velocity of the material in the wake as a function of distance from the star observed by \citet{Mathews2008}. Interestingly, \citet{Raga2008} predict a broader tail than what is observed, with an opening angle of $24^\circ$. This is partially because their model assumes efficient mixing, partially because the part of the tail that is detected in the UV is only the inner core. As the right panel of Figure~\ref{fig:Mira2_slice} shows, the wind material, traced by the tracer fluid, indeed spreads out with an opening angle, but only the densest part in the center can be observed.

\citet{Martin2007} report evidence for a cyclic behavior in Mira's UV tail. Their periodogram analysis suggests a period of $10^4$ yr and a secondary peak at about half this value. Given the velocity they assume for the tail ($\rm \sim 130\, km/s$), this translates to periods of $\sim 1.3$ pc and $\sim 0.6$ pc. We have carried out a Lomb-Scargle Periodogram analysis of our simulated tail of Mira with cooling (upper right panel of Figure~\ref{fig:2by2}), and found a peak at about $\sim 0.5$ pc. The result is the same for all snapshots in the steady state. This period is consistent with the secondary peak found in \citet{Martin2007}, and its origin is the turbulent interaction between Mira's wind and the ISM. The first peak may be due to physics we have not included (e.g., thermal pulses or magnetic fields). Note that \citet{Martin2007} rule out the possibility of thermal pulse because theoretical models predict thermal pulse cycles at least an order of magnitude longer than their estimated periodicity timescale. However, their timescale estimation is based on an assumed velocity of the tail that is an order of a magnitude too large (in the rest frame of the star) (see Section~\ref{sec:previous1} for more discussion). Therefore the timescale of the period is under-estimated by an order of magnitude, and should be on the order of $10^5$ yr. This makes thermal pulse a plausible explanation for the observed periodicity in Mira's tail.

\subsection{Implications for Elliptical Galaxies and Galaxy Clusters}\label{sec:dust}

Most cool-core galaxy clusters, as well as a significant fraction of giant elliptical galaxies, harbor extended multiphase gas in the center \citep{McDonald10, Werner2014, Pandya2017}. These multiphase structures often appear filamentary on small scales. These filaments can be observed via the H$\alpha$ and other emission lines. Numerical studies suggest that the cold filaments may condense out of the hot gas due to thermal instability \citep{McCourt12, Sharma2012}. Simulations of idealized galaxy clusters and elliptical galaxies show that kinetic AGN feedback can trigger local instabilities via uplifting lower-entropy (but still hot) gas \citep{PII, Voit2017} while globally suppressing cooling \citep{PIII, Wang2018}. These precipitation-regulated feedback models have achieved great success in generally reproducing many observed properties of massive galaxies and galaxy clusters, including the level of cooling (X-ray luminosity), the velocity dispersion, the star formation rate and the general morphology of the cold, star forming structures \citep{Gaspari2012, Li2015, Prasad2015}. 

However, two problems remain with these models that form cold gas from thermal instabilities. First, as pointed out by \citet{Hogan2017}, the minimum cooling to free-fall time ratio $t_{cool}/t_{ff}$ tends to be higher in the observations than in simulations \citep{Li2015, Prasad2018}. In particular, cool-core clusters in idealized simulations sometimes have $min(t_{cool}/t_{ff}) < 10$, which is rarely seen in real clusters. One possible explanation for this discrepancy is that minor mergers, which are not included in these idealized simulations, may trigger condensation before $min(t_{cool}/t_{ff})$ drops much below 10. However, this has not been tested in numerical simulations. 

The second problem relates to the origin of dust in these cold filaments. The observed multi-phase filaments are dusty with polycyclic aromatic hydrocarbons (PAHs) \citep{Donahue2011}. Dust grains can form in the winds of AGB stars and during Supernova explosions. Type II supernovae (SNeII) are rare in early-type galaxies. Type Ia Supernovae (SN Ia) are also less abundant than AGB stars, and \citet{Nozawa2011} suggest that the newly formed dust in SN Ia is almost completely destroyed in the shocked gas before being injected into the interstellar medium. Therefore, we mainly consider dust produced in AGB winds. 

In hot environments such as the centers of galaxy clusters and massive elliptical galaxies, dust grains can be quickly destroyed by high energy ions through thermal sputtering. Following \citet{Tsai1995} and \citet{McKinnon2017}, the sputtering timescale of dust grains can be estimated as 

\begin{equation}
t_{sp} = (0.17 \rm \, Gyr) \Big(\frac{a/0.1\mu m}{\rho/10^{-27}g\,cm^{-3}}\Big) \Big[ \Big(\frac{2\times 10^6K}{T}\Big)^{2.5} + 1\Big] \, ,
\end{equation}
where a is the grain size and $\rho$ and $T$ are the ambient gas density and temperature. The cooling time of the gas is 

\begin{equation}
t_{cool} = \frac{5nk_BT}{2n^2\Lambda(T)} = \frac{5k_B T\mu m_p}{2\rho \Lambda(T)} \, ,
\end{equation}
where $k_B$ is the Boltzman constant and $\Lambda (T)$ is the cooling function. 

\begin{figure}
\begin{center}
\includegraphics[scale=0.4,trim=0cm 0cm 0cm 0cm, clip=true]{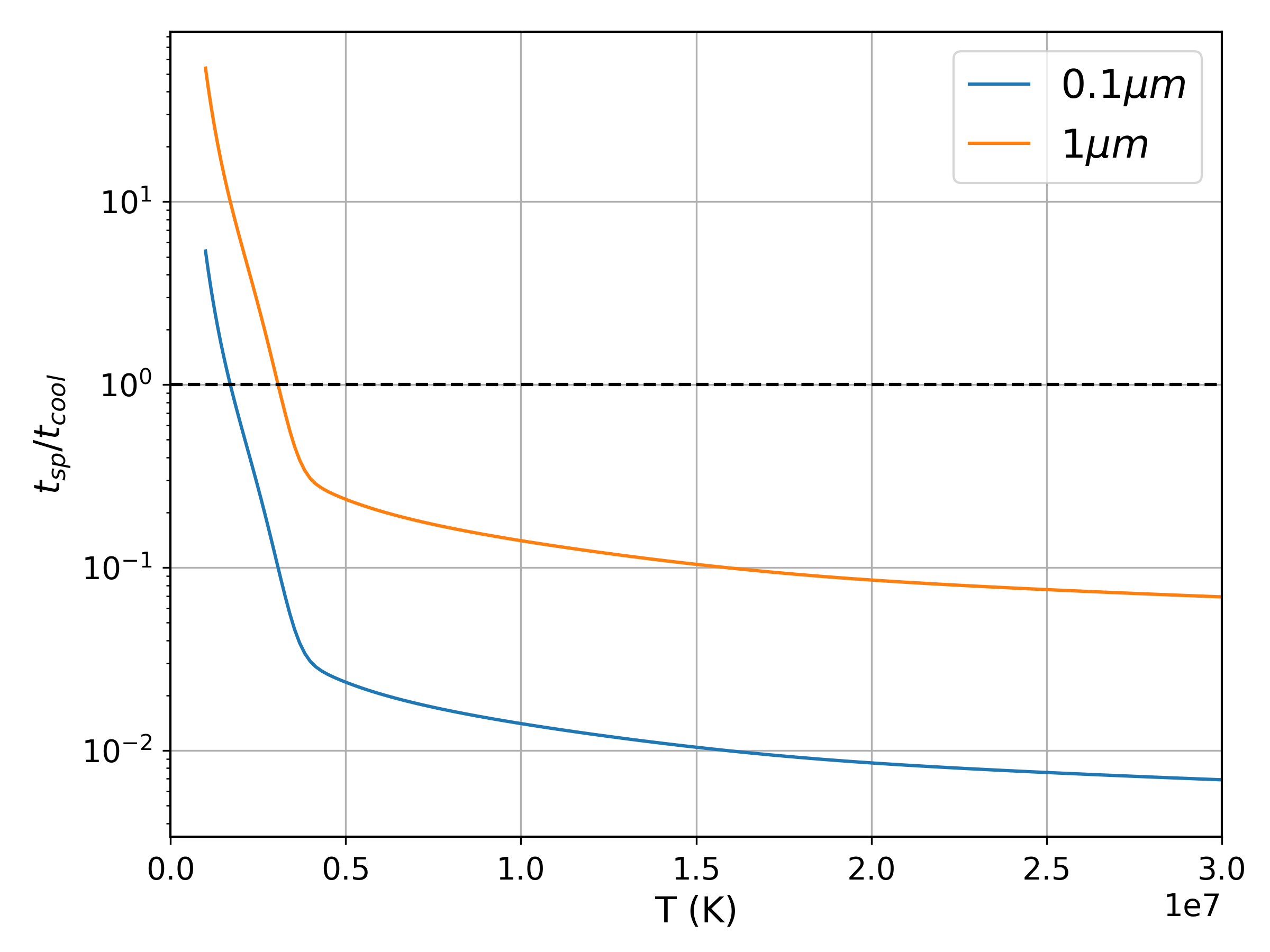}
\caption{The sputtering time of dust grains divided by the cooling time of the gas as a function of temperature. The blue line corresponds to typical interstellar dust grains of $0.1\mu m$. The orange line corresponds to grain size of $1\mu m$, which is slightly larger than the upper limit of grain size distribution derived from observations \citep{Weingartner2001}. The sputtering-to-cooling time ratio drops below 1 (black dashed line) even for the largest grains at temperatures above $\sim 3\times 10^6$ K.
\label{fig:sputtering}}
\end{center}
\end{figure}

Thus the ratio between sputtering time and cooling time is independent of gas density. 
Figure~\ref{fig:sputtering} shows the sputtering-to-cooling time ratio as a function of temperature for two different grain sizes. The blue line is for typical interstellar dust grains of $0.1\mu m$. The orange line corresponds to grain size of $1\mu m$. The grain size distribution in most dust models has a sharp cut-off at large size, and it is always smaller than $1\mu m$ \citep[e.g.,][]{Weingartner2001, Draine2007}. Thus the orange line can be treated as an absolute upper limit. Sputtering time of dust grains in the hot gas in the center of galaxy clusters is extremely short ($\sim 10$ Myr), which is also recently shown in numerical simulations \citep[e.g.,][]{Vogelsberger2018}. The ICM is essentially dust-free \citep{Fabian1982}. 

If AGB winds mix quickly with the hot surrounding gas, and if cold gas forms out of cooling instability in the dust-free ICM, in order for the cold gas to be dusty, the sputtering time has to be longer than the cooling time. As Figure~\ref{fig:sputtering} shows, this is not the case even for the largest grains in the center of massive galaxies and galaxy clusters with temperatures higher than $\sim 3\times 10^6$ K. All the dust grains should be sputtered before the hot ICM precipitates into cold filaments, and therefore the filaments should be dust-free. If the dust grains are only partially destroyed, one would expect the dust size distribution in the filaments to be biased towards large grains. However, observations of cold filaments suggest that the dust and PAHs are similar in their size distributions as in normal spiral galaxies \citep{Donahue2011}, and therefore they must have been shielded from the hot ICM. 

One solution to this problem is assuming that AGB winds do not fully mix with the surrounding hot gas. As our simulations show, in high pressure environments such as the centers of massive galaxies, the cold AGB wind can survive longer than adiabatic mixing as a result of efficient cooling of the mixed material. In one scenario, the dusty cold gas survives in the form of a mist \citep{McCourt2018, Liang2018} which co-moves with the hot gas, and when cooling instability happens, these dusty cloudlets join the newly cooled gas and pollute it with dust. This will only work if the small cloudlets are magnetically isolated from the hot gas, since otherwise they will evaporate very quickly due to conduction. Instead, we speculate that more of the mixed gas can cool further down the stream and result in a higher cold gas mass than what is originally in the AGB wind. This process of condensation due to mixing is analytically described in \citet{Begelman1990} and recently shown numerically in \citet{GronkeOh2018}. 

In this scenario, the origin of multiphase gas in the center of massive systems is AGB wind induced cooling, rather than thermal instability. Condensation (or precipitation) still happens, but it happens at the turbulent mixing layer between hot and cold phases. This model naturally explains the existence of dust and PAHs in the cold filaments\footnote{We do not claim that all the dust found in cold filaments is from AGB winds. We expect that there is growth of dust in the cold filaments as in the ISM \citep{Draine2009}. AGB wind is only responsible for the seeds, but not the total dust mass.}. Moreover, $min(t_{cool}/t_{ff})$ of the hot gas is likely higher when condensation due to mixing happens, and thus agrees with the observations better. The correlation between the existence of multiphase gas and short $t_{cool}$ of the hot gas will still exist because the shorter $t_{cool}$ gives a shorter cooling time of the mixed gas. This model is promising, but is certainly incomplete. We discuss the limitations of this study and future directions in Section~\ref{sec:last}.

\subsection{Comparison with Previous Works}\label{sec:previous}
In this section, we compare our results with previous simulation works that are relevant. 
\subsubsection{Comparison with Previous Mira Simulations}\label{sec:previous1}
Previous simulations of Mira's tail have been mostly focused on explaining its observed appearance, in particular, the shape of the tail. To explain the broad-head/narrow-tail structure, both \citet{Wareing2007} and \citet{Esquivel2010} evoke the scenario that Mira has only recently entered the Local Bubble. \citet{Wareing2012} further assumes that Mira entered the Local Bubble at an angle to explain the observed kink of the tail. Since Mira itself is not the focus of this work, we instead use a simpler setup assuming Mira has always been in the Local Bubble, similar to \citet{MiraMHD2013}, which allows us to focus on the effects of mixing and cooling. Even without fine-tuning Mira's environment or trajectory, our simulated tail of Mira already bears remarkable resemblance to the observed tail. 

In spite of the difference in the simulation setup and the focus of the study, we find that our results are generally consistent with previous simulations of Mira. Like previous works \citep{Wareing2007,Esquivel2010}, we find that the tail of Mira is highly turbulent. The formation time of Mira's tail in our simulation is $\sim 200$ kyr (see Figure~\ref{fig:Mira2_cold_flux}). This is comparable to the estimates in previous simulations of \citet{Wareing2007} ($\sim 450$ kyr) and \citet{Esquivel2010} ($\sim 300$ kyr), and is much longer than the estimates assuming that the wind material is instantly decelerated to zero velocity \citep{Martin2007} ($\sim$ 30 kyr).

\subsubsection{Comparison with \citet{Parriott2008}}
\citet{Parriott2008} use 2D hydrodynamic simulations to study the fate of AGB winds in early-type galaxies. Our LP simulation uses the same parameters as their fiducial run, where they find that $\sim 25\%$ of the cold wind survives and leaves the simulation box ($\sim 25$ pc from the star) in a laminar flow. In our LP run, the tail is fully mixed at less than 20 pc from the star. We have verified that this is not due to the axisymmetric setup in \citet{Parriott2008}, as our results for the LP run stay roughly the same when we use an axisymmetric setup and only simulate half of the star. The resolution of the two simulations is also similar. We have lowered our resolution to match that of \citet{Parriott2008} and again found a fully mixed tail in the LP run. Thus the difference is likely because 2D and 3D simulations can produce different results, as seen in other simulations in different contexts \citep{Zhang2004}. The behavior of turbulence can be very different in 2D due to inverse cascade. Instabilities can be suppressed in 2D, as is recently shown in simulations by \citet{Mandelker2018} and \citet{Sparre2018}. This explains why \citet{Parriott2008} is able to form a laminar flow which is not seen in our 3D simulation. 

Similarly to our work, \citet{Parriott2008} also find little difference between the adiabatic run and the cooling run with their fiducial setup as cooling is not important in this regime. In their parameter studies, \citet{Parriott2008} find that when the ambient density is lower, less cold gas survives. This is in agreement with our analytic prediction in Section~\ref{sec:analytic} and the trend we see in our simulations.

\subsection{Caveats, Limitations and Future Directions}\label{sec:last}

\begin{figure}
\begin{center}
\includegraphics[scale=0.3,trim=0cm 0cm 0cm 0cm, clip=true]{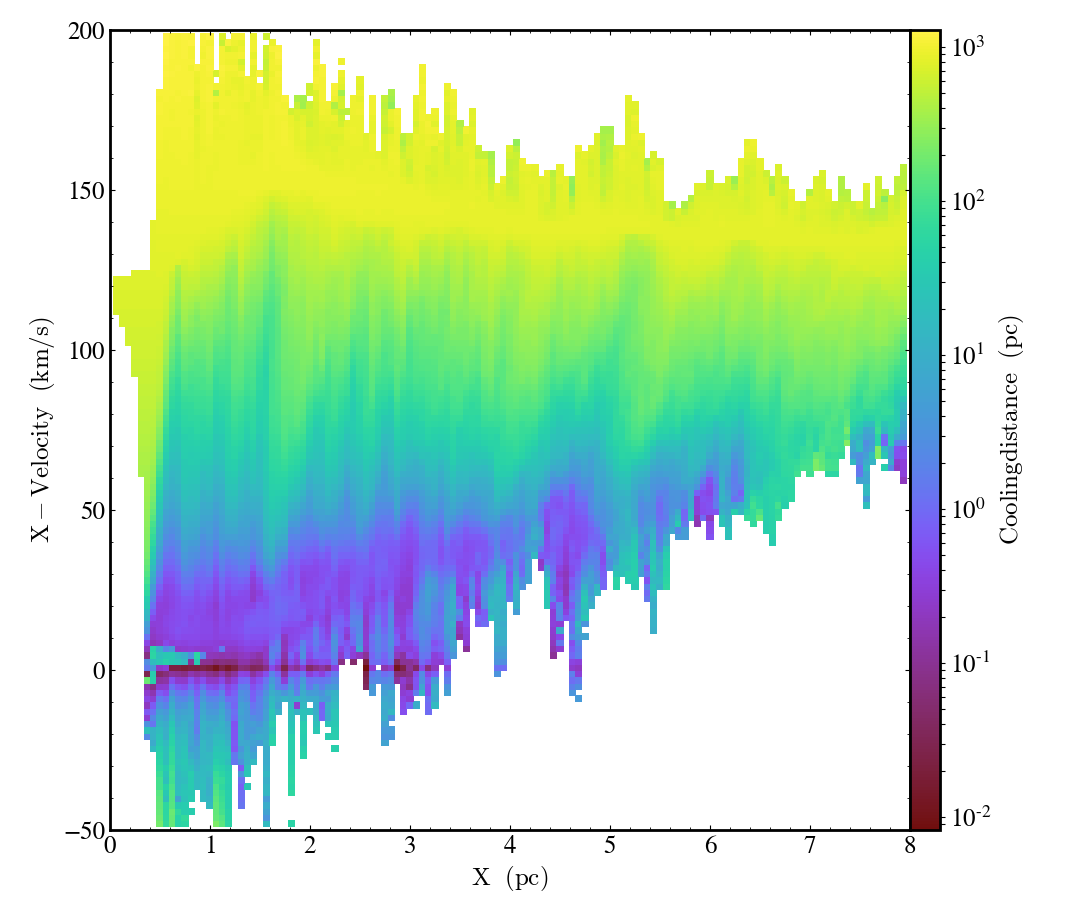}
\caption{The ``cooling distance'' as a function of position and x-velocity. Cooling distance is defined as the distance the gas travels within a cooling time. A significant fraction of the mixed gas will move far outside the simulation box before it can cool. 
\label{fig:cooling_distance}}
\end{center}
\end{figure}

One major caveat of our simulations, as mentioned previously, is that we are not modeling the time dependent AGB wind correctly. In order to understand the basic physical processes of mixing and cooling, we use a simple assumption that the mass-loss rate is a constant. This assumption is also used in previous numerical studies of Mira and AGB winds \citep{Wareing2007, Parriott2008}. However, the mass-loss rate, as well as the velocity of the stellar wind, is a function of time, and increases drastically towards the end of the AGB phase when planetary nebula is formed. The planetary nebula ejecta contains a significant fraction of the mass return and possibly contributes more to induced cooling than the earlier AGB phase \citep{Bregman2009}. Our future study will use a time-dependent mass-loss rate based on stellar evolution models \citep{MIST} and follow the entire AGB phase through the formation of planetary nebula. 

Another caveat is that our box is not long enough to follow the later evolution of the mixed stellar wind and ISM. Figure~\ref{fig:cooling_distance} shows the ``cooling distance'', the distance gas travels within its cooling time, as a function of position and x-velocity. The high velocity gas is the hot ISM wind, and the gas with the lowest positive velocity is the accelerating stellar ejecta. Because the gas is highly turbulent, some of it shows negative x-velocity. A significant fraction of the mixed material has a cooling distance much larger than the box size. As \citet{GronkeOh2018} show, cooling of the mixed material can happen later at larger distances, and this is not captured in simulations with boxes not long enough. 

We have also only explored a small range of the parameter space, in terms of the properties of the AGB wind, the velocity of the star, and the physical condition of the ambient environment. Our analytical calculations and simulations suggest that cooling is more important in higher pressure environments. We expect induced cooling to be even stronger in the center of a cool-core cluster than our HP run or Mira. Unfortunately we could not complete such a simulation because higher pressure corresponds to a smaller stand-off radius, and thus requires higher resolution. With our current simulation setup, a cluster run is computationally too expensive. We thus defer a cluster run to future studies where we model the planetary nebula formation phase and the later interaction with the hot ICM separately.

Lastly, the simulations in this work do not include some physical processes that may be important, such as conduction, magnetic fields and dust grains. Magnetic fields have been shown to prolong the survival time of cold clouds agains hot wind \citep{McCourt2015}, and conduction can evaporate cold gas, shortening its survival time \citep{Liang2018}. We defer the study of the effects of conduction and magnetic fields to future works. However, the fact that our simulated Mira's tail appears similar to the real tail suggests that conduction is likely suppressed. It is still possible that magnetic fields are important, and Mira's tail is still growing and has not reached a steady state yet. Dust grains can be an important coolant \citep{Draine1981}. Including dust cooling will likely result in more efficient cooling in the mixing layer.

\section{Conclusions}\label{sec:conclusions}
In this work, we have studied the properties of the tail of an AGB star moving through hot gas using both analytical calculations and numerical simulations. We performed three sets of 3D hydro simulations of AGB winds interacting with different environments: the Local Bubble (Mira), outskirts of elliptical galaxies, central regions of massive elliptical galaxies. The key findings are summarized below.

\begin{enumerate}
\item The wind from a fast moving AGB star forms a trailing tail due to the ram pressure of the surrounding medium. In the absence of cooling, we show analytically that the head-to-tail ratio should be about $1/4$ for all AGB stars in such a configuration regardless of their environment. Our three adiabatic simulations of three AGB stars in different environments confirm this finding.

\item For a typical AGB star moving at a typical stellar velocity in elliptical galaxies, analytically, the cooling time of the tail is inversely proportional to the ambient pressure, suggesting that cooling is more important in higher pressure environment. 

\item When we compare our simulations of an AGB star in the outskirts of elliptical galaxy with and without radiative cooling, we find little difference. This is because cooling is not important, and mixing is the only important physical process. Unlike previous 2D simulations, we find no cold laminar flow leaving the simulation box. All the stellar ejecta are fully mixed with the hot surrounding gas.

\item With radiative cooling, the tail of Mira is more than twice as long as the adiabatic case, and the length is consistent with the observed tail of Mira ($\sim 4$ pc). The lengthening of the tail is because some of the gas in the mixing layer between the AGB wind and the hot ISM cools efficiently. The tail of an AGB star in the central regions of elliptical galaxies is very similar to the tail of Mira, with a lengthened tail due to induced cooling. 

\item The velocity of the simulated tail of Mira is consistent with that measured from the HI observations of Mira. The formation time of Mira's tail in our simulation is $\sim 200$ kyr. \citet{Martin2007} finds a period of  $\sim 1.3$ pc and $\sim 0.6$ pc in Mira's UV tail. Our simulated Mira's tail has a periodicity consistent with the second one. We suggest that the first period may be due to thermal pulses of Mira A.

\item The dust sputtering time is shorter than the cooling time of the hot gas in the central regions of massive galaxies and galaxy clusters. Many of these systems host cold filaments that are observed to be dusty with PAH molecules. If the cold filaments form due to thermal instability, they should be dust free. We speculate that the dusty cold filaments form as a result of induced cooling in the mixing layer between dusty AGB winds and the surrounding hot gas. This is the most plausible explanation for the existence of dust in the cold gas.

\end{enumerate}
 
Our analytical and numerical work shows that in high pressure environments (e.g., Local Bubble, central regions of elliptical galaxies and galaxy clusters), the interaction between AGB wind and the hot ISM can introduce condensation in the mixing layer. This explains the long comet-like tail of Mira. We propose an alternative model for the origin of cold dusty filaments in massive systems -- instead of condensation due to thermal instability of the hot medium, condensation happens due to mixing of dusty AGB wind and hot gas. Future studies should use a more realistic model of stellar mass loss rate as a function of time and investigate the interaction between a planetary nebula and its surrounding hot medium. The effects of magnetic fields and thermal conduction should also be examined in future simulations.

\section*{Acknowledgements}

Computations were performed using the publicly-available Enzo code, which is the product of a collaborative effort of many independent scientists from numerous institutions around the world. Their commitment to open science has helped make this work possible. Data analysis and visualization are partly done using the \textsf{yt} project \citep{yt}. The simulations are performed on the Rusty cluster of the Simons Foundation. We thank the Scientific Computing Core of the Simons Foundation for their technical support. This work was supported in part by a Simons Investigator Award from the Simons Foundation, by NSF grant AST-1715070, and by Grant 528306 from the Simons Foundation. GLB was partially supported by NSF grant AST-1615955 and NASA grant NNX15AB20G.

\vspace{0.2in}

%\bibliography{library}

\end{document}